\documentclass[journal=jacsat,manuscript=article]{achemso}
\pdfoutput=1

\usepackage[version=3]{mhchem} 
\usepackage[utf8]{inputenc} 
\usepackage[T1]{fontenc} 
\usepackage{graphicx}       
\usepackage{amsmath, amssymb}        
\usepackage{breakcites}     
\usepackage{array}          
\usepackage{multirow}       
\usepackage{threeparttable}
\usepackage{pdflscape}
\usepackage{xcolor}
\usepackage{bibunits}
\usepackage{lscape}
\usepackage{tablefootnote}
\usepackage{longtable}
\SectionNumbersOn

\graphicspath{ {./images/} }

\author{Thijs Stuyver}
\affiliation[MIT]
{Department of Chemical Engineering, Massachusetts Institute of Technology, 77 Massachusetts Avenue, Cambridge, Massachusetts 02139, United States}
\author{Connor W. Coley}
\affiliation[MIT]
{Department of Chemical Engineering, Massachusetts Institute of Technology, 77 Massachusetts Avenue, Cambridge, Massachusetts 02139, United States}
\altaffiliation{Department of Electrical Engineering and Computer Science, Massachusetts Institute of Technology, 77 Massachusetts Avenue, Cambridge, Massachusetts 02139, United States}
\email{ccoley@mit.edu}

\title{Machine learning-guided computational screening of new bio-orthogonal click reactions}

\abbreviations{}

\begin{document}


\begin{abstract}
Bio-orthogonal click chemistry has become an indispensable part of the biochemist's toolbox. Despite the wide variety of applications that have been developed in recent years, only a limited number of bio-orthogonal click reactions have been discovered so far, most of them based on (substituted) azides. In this work, we present a computational workflow to discover new candidate bio-orthogonal click reactions. Sampling only around 0.05\% of an overall search space of over 10,000,000 dipolar cycloadditions, we develop a machine learning model able to predict DFT-computed activation and reaction energies within $\sim$2-3 kcal/mol across the entire space. Applying this model to screen the full search space through iterative rounds of learning, we identify a broad pool of candidate reactions with rich structural diversity, which can be used as a starting point or source of inspiration for future experimental development of both azide-based and non-azide-based bio-orthogonal click reactions.
\end{abstract}
\begin{bibunit}[unsrt]

\section{Introduction}\label{sec:intro}
Bio-orthogonal click chemistry -- and the revolutionary capabilities it offers in terms of real-time, \emph{in vivo} imaging and tracking -- has had a profound impact on biochemical research. \cite{saxon2000cell, hang2003metabolic, agard2004strain, prescher2004chemical, prescher2005chemistry, hackenberger2008chemoselective, jewett2010cu, sletten2009bioorthogonal} Bio-orthogonal reactions have been employed to detect, track and study individual biomolecules, e.g., (m)RNA, metabolites etc., as well as to label/tag entire (tumor) cells. Furthermore, many advanced applications related to medicine and healthcare, e.g., targeted cancer treatments and \emph{in situ} drug assembly, are currently under development. \cite{nwe2009growing,devaraj2018future, arkenberg2020recent}

To be considered a suitable bio-orthogonal click reaction, reactions need to fulfill a broad range of criteria: (a) the reactants should be sufficiently inert so that they do not degrade rapidly under physiological conditions, (b) side reactions between the reactants and native biomolecules should not occur at appreciable rates to ensure adequate selectivity, (c) the rate of the reaction should be sufficiently high at room/body temperature so that ligation takes place at a faster timescale than clearance of the reactants from the studied organism, and (d) the reactive components of the reactants should be able to be integrated into biomolecules through metabolic or protein engineering. Ideally, the compounds will be sufficiently small so as to not disturb the native behavior of the biological system under investigation. \cite{jewett2010cu}

\begin{figure}
\centering
\includegraphics[scale=0.48]{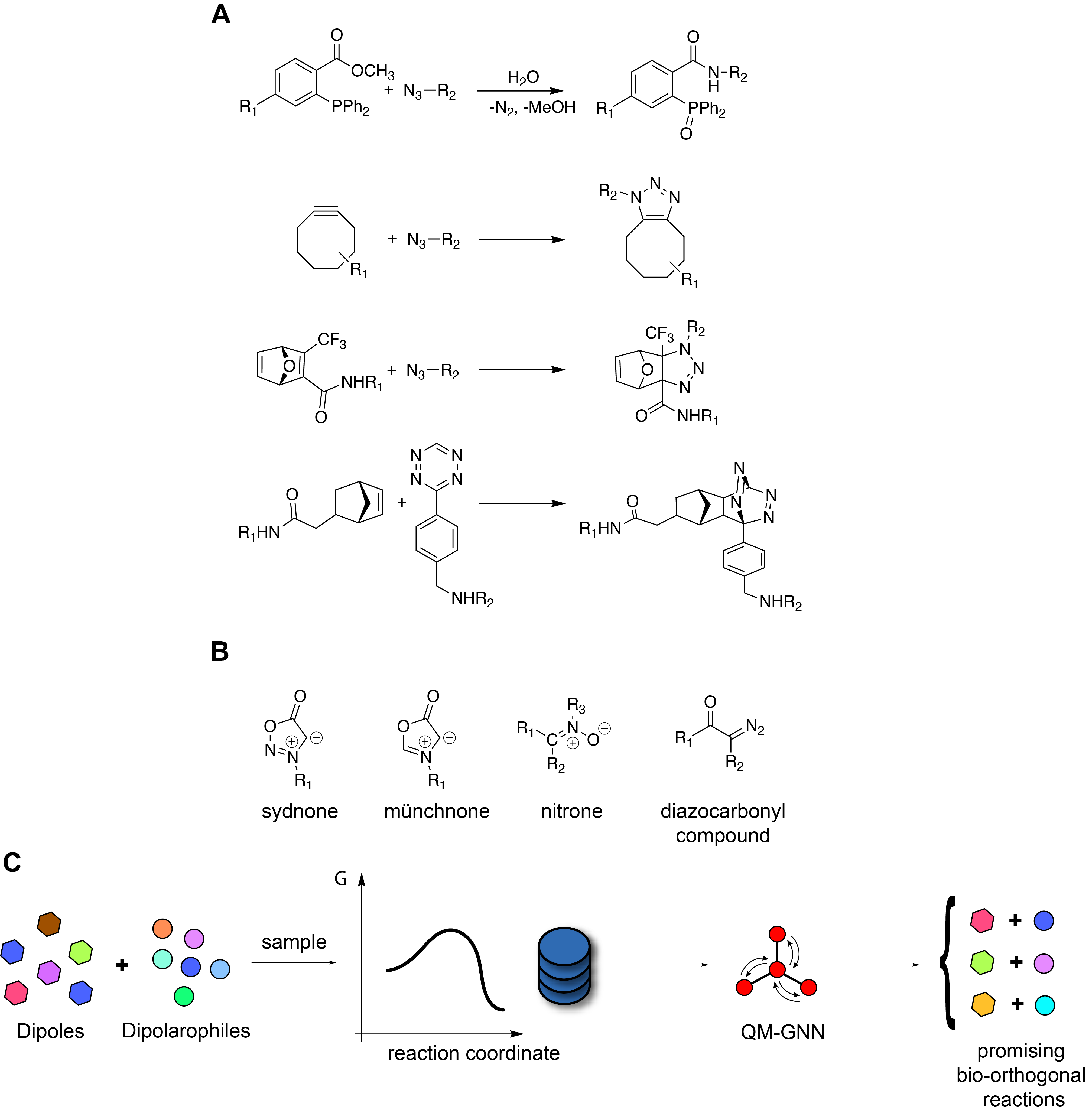}
\caption{(A) The Staudinger ligation reaction \cite{saxon2000cell, prescher2004chemical} (top), and some popular strain-based cycloaddition reaction scaffolds involving cyclooctyne, \cite{agard2004strain} oxo-norbornadiene \cite{van2007metal} and norbornene \cite{devaraj2008tetrazine} (bottom). (B) Examples of alternative dipoles that have been considered for bio-orthogonal click applications\cite{breugst2020huisgen, porte2020click, narayanam2016discovery}. (C) Overview of the workflow taken in this work. First, a set of dipoles and dipolarophiles are generated. Next, a sample of combinations is taken and reaction profiles are computed for each of them. The resulting dataset is then used to train a QM-augmented graph neural network (QM-GNN) model, which finally enables a high-throughput screening of the full chemical space and identification of promising bio-orthogonal reactions.}
\label{fig:bio_ortho_examples}
\end{figure}

These many requirements which need to be fulfilled make the search for new bio-orthogonal click reactions reminiscent of looking for a needle in a haystack. Consequently, only a fairly limited number of such reactions have been identified so far. In their seminal work in which the concept of bio-orthogonal click chemistry was introduced, Bertozzi and co-workers took inspiration from the Staudinger ligation reaction between azides and triarylphosphines equipped with an ester. \cite{saxon2000cell, prescher2004chemical} Following this first successful demonstration, the focus within the field has to a significant extent remained on (substituted) azides: many -- though definitely not all \cite{kim2015bioorthogonal, sletten2011bioorthogonal, nguyen2020developing,blackman2008tetrazine, oliveira2017inverse} -- of the subsequently developed bio-orthogonal reactions also involve (derivatives of) this specific 1,3-dipole (see Figure \ref{fig:bio_ortho_examples}A for some representative examples). \cite{liu2019bio, jewett2010cu}

Particularly notable in this regard are the [3+2] cycloaddition reactions between azides and strained dipolarophiles such as cyclooctyne \cite{agard2004strain} and oxo-norbornadiene \cite{van2007metal} which have been applied with great success in diverse settings over the past two decades. \cite{jewett2010cu} Attempts have been made to expand the scope of bio-orthogonal cycloaddition reactions to other types of dipoles (and dipolarophiles), e.g., nitrones \cite{dommerholt2010readily, bilodeau2021bioorthogonal}, diazocarbonyls \cite{andersen2015diazo} and cyclic dipoles such as sydnones and m\"unchnones \cite{breugst2020huisgen, porte2020click, narayanam2016discovery, decuypere2017sydnone} (Figure \ref{fig:bio_ortho_examples}B), as well as to various types of inverse-electron demand Diels-Alder reactions, \cite{devaraj2008tetrazine, blackman2008tetrazine, ravasco2020predictive} and specific advantages over azide-based reactions have been claimed/demonstrated for some of these alternatives (e.g., reduced steric bulk \cite{andersen2015diazo}, or improved kinetic properties \cite{svatunek2022uncovering, oliveira2017inverse}) but nonetheless, the azide-based reactions have maintained their popularity in bio-orthogonal click applications. \cite{breugst2020huisgen}

\begin{figure}
\centering
\includegraphics[scale=0.35]{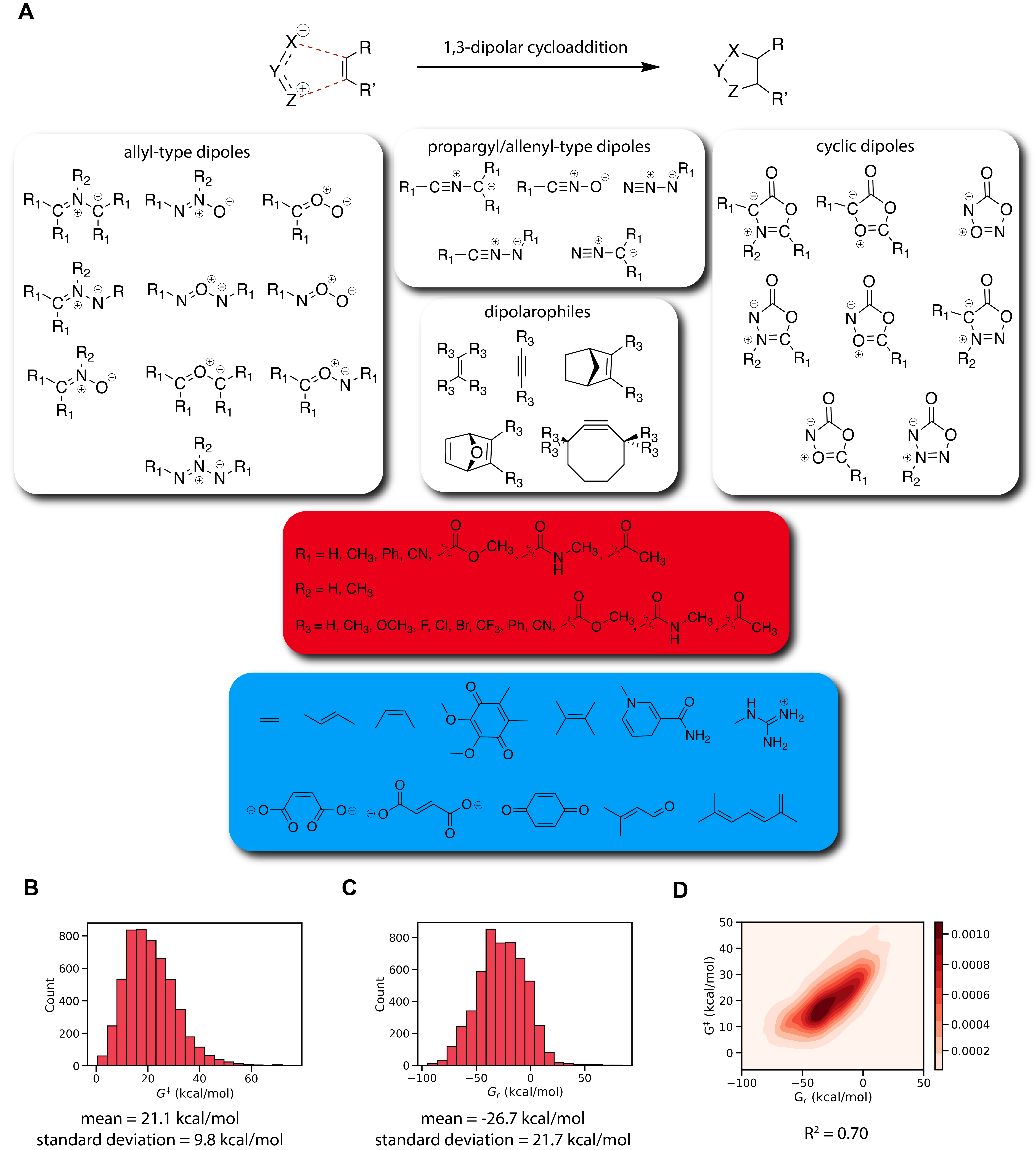}
\caption{(A) Schematic overview of the search space for the considered dipoles and dipolarophiles, with the synthetic dipoles and dipolarophiles at the top, the substituents in the red box and the biologically relevant fragments in the blue box. The synthetic dipolarophile scaffolds are respectively ethene, ethyne, norbornene, oxo-norbornadiene and cyclooctyne.  (B) Histogram representing the distribution of the computed activation energies (G$^{\ddagger}$). (C) Histogram representing the distribution of the computed reaction energies (G$_{r}$). (D) Scatter density plot showing the correlation between activation and reaction energies.}
\label{fig:overview}
\end{figure}

In this study, we make use of our recently published computational dataset of [3 + 2] cycloaddition reactions \cite{stuyver2022reaction}  to train data-driven models which can be employed to screen for new, promising candidates for bio-orthogonal click reactions (Figure~\ref{fig:bio_ortho_examples}C). More specifically, we train an ensembled multitask QM-augmented neural network to simultaneously predict activation and reaction energies for every combination within the broad chemical space spanned by the dataset, comprising 3555 dipoles, 1339 synthetic dipolarophiles, as well as 12 biologically relevant motifs which can act as alternative/competing dipolarophiles (cf. Figure \ref{fig:overview} for an overview of the full chemical space probed). Exhaustively enumerating all possibilities, almost 5M potential reacting systems (and 10M+ unique reactions when regiochemistry is taken into account) are contained within this space. Computing reaction profiles for every point in this chemical space following our protocol in ref. \citenum{stuyver2022reaction} would require approximately 2 billion CPU hours, or 230.000 CPU years. Therefore, we use only a sparse sample and employ machine learning techniques to facilitate efficient model-guided exploration of the full space.

As will be demonstrated below, the candidate pool resulting from our screening campaign contains a rich structural diversity, which can be used as a starting point/source of inspiration for future experimental development campaigns of both azide- and non-azide-based bio-orthogonal click reactions.

\section{Methodology}

\subsection{Dataset}

As input for (the "zeroth" iteration) of our screening campaign, we use our recently published dataset. \cite{stuyver2022reaction} This dataset consists of 5271 computed reaction profiles for [3 + 2] cycloaddition reactions, sampled from the chemical space outlined in Figure \ref{fig:overview}, at B3LYP-D3(BJ)/def2-TZVP//B3LYP-D3(BJ)/def2-SVP, \cite{becke1988density, lee1988development, stephens1994ab, grimme2006semiempirical, schafer1992fully, schafer1994fully} with standard conditions and an (aqueous) SMD model \cite{marenich2009universal} imposed to mimic physiological conditions. For more information about the methodology, the benchmarking of the level of theory and the validation of the approach, as well as the sampling strategy of the chemical space, we refer readers to the original publication. 

In the version used in this work, two data points with incorrect reaction energy values -- which had not been identified as such by our original postprocessing workflow -- were still present. These incorrect data points were filtered out after the first iteration of our active learning procedure (\emph{vide infra}), and have also been corrected in ref.~\citenum{stuyver2022reaction}.

The validation of model-predicted hits, and hence the acquisition of new training data in subsequent iterations of the active learning procedure, were performed according to the same computational methodology.

\subsection{Model architecture}

A graph neural network (GNN), consisting of a combination of a Weisfeiler-Lehman (WL) convolutional embedder and a global attention mechanism, was selected as the machine learning model of choice (cf. Section \ref{sec:architecture} for a more in-depth discussion of the individual network branches). Other -- more simple -- model architectures based on both fingerprints and reactivity descriptors extracted from the reactants (\emph{vide infra}), were tested as well (e.g., multivariate linear regression, component-based k-nearest neighbors, random forests, XGBoost, etc.), but none of them reached a comparable performance to the GNN (Sections  \ref{sec:bayesian_opt} and \ref{sec:alternative_algorithms}).

\begin{figure}
\centering
\includegraphics[scale=1]{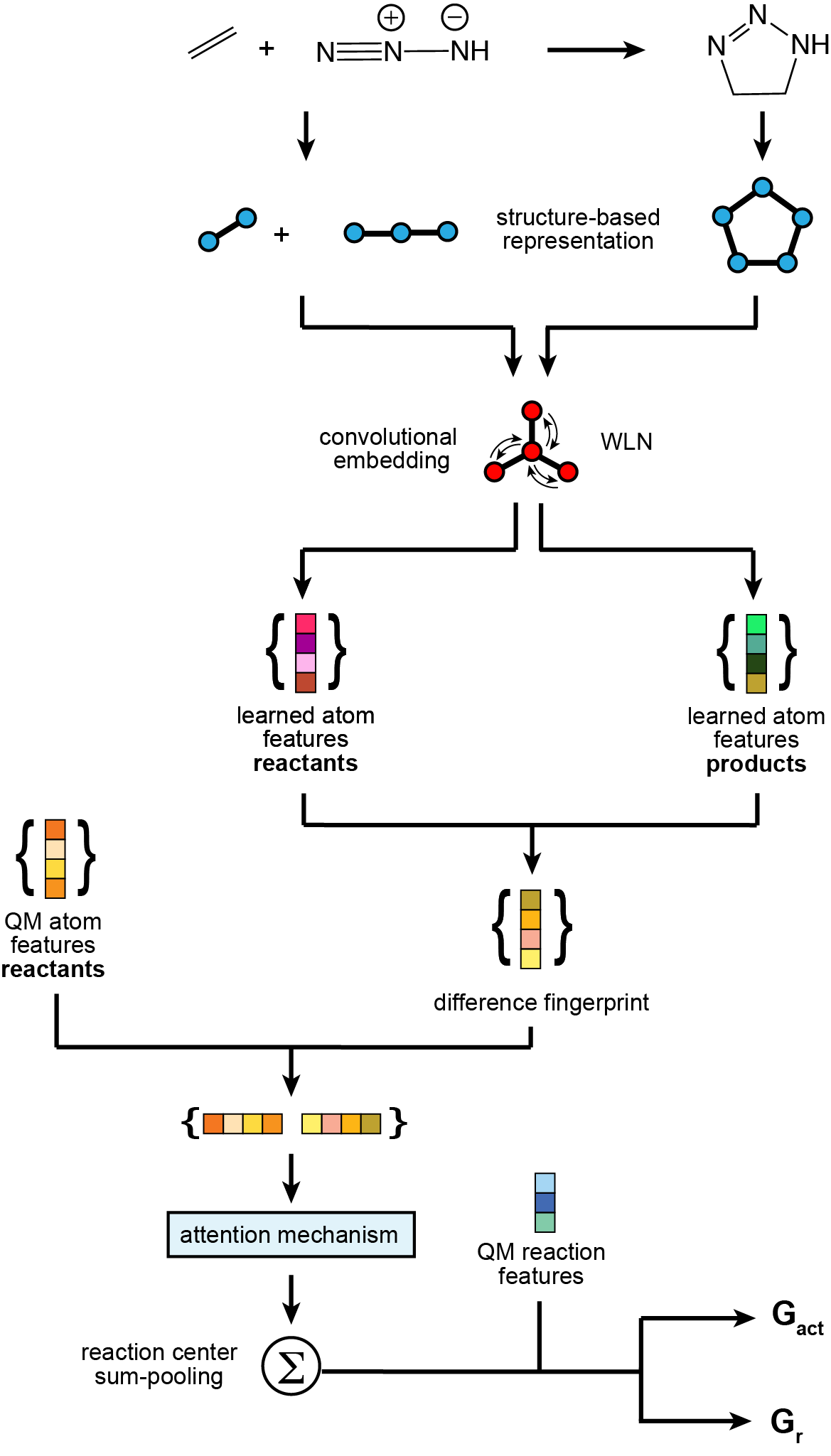}
\caption{A schematic overview of the ml-QM-GNN model architecture. WLN denotes the Weisfeiler-Lehman network branch.}
\label{fig:model_architecture}
\end{figure}

A schematic overview of the GNN model is provided in Figure \ref{fig:model_architecture}. As input, the model takes the simplified molecular-input line-entry system (SMILES) representations of both reactants and products, parses them into graph-based representations using RDKit,\cite{landrum2006rdkit} and calculates structural descriptors (atomic number, formal charge, ring status, bond order, etc.) for each heavy atom and bond. The resulting structure-based representations for reactants and products are subsequently passed through the WL embedder, after which the difference fingerprint is taken, i.e., the product representation is subtracted from the reactant representation. The set of atomic feature differences is then passed through a global attention mechanism \cite{yang2016hierarchical} to capture the influence of distant parts of the reacting system, followed by another dense feedforward layer. Next, the atom-level representations are sum-pooled into a single molecule-level representation which yields, after a couple more dense feedforward layers, the predicted values for the activation and reaction energies. A Bayesian Optimization-based hyperparameter tuning was performed with hyperopt \cite{bergstra2015hyperopt} to select a reasonable set of hyperparameters (Section \ref{sec:bayesian_opt}).

In line with previous work within our research group, \cite{stuyver2022quantum, guan2021regio} we introduced QM-augmentation by enriching the atom- and molecule/reaction-level featurization with quantum chemistry-motivated reactivity descriptors from the reactants, respectively atomic charges, NMR shielding constants, nucleophilic/electrophilic Fukui functions, \cite{murray2011electrostatic,geerlings2003conceptual} and Valence Bond reactivity theory derived promotion gaps \cite{shaik2007chemist, stuyver2021promotion}. Definitions of and motivations for inclusion of our set of descriptors are provided in Section \ref{sec:QM_descriptor_motivation}. It is worth mentioning here that in a recent study by Coelho and co-workers, quantum chemistry-based descriptors were demonstrated to facilitate the construction of accurate predictive models for small datasets of experimental rate constants for bio-orthogonal inverse-electron demand Diels-Alder reactions. \cite{ravasco2020predictive}  

All descriptors were explicitly computed here at B3LYP/def2-SVP//GFN2-xTB level of theory for each dipole and dipolarophile present in the dataset according to the previously developed in-house workflow outlined in Section \ref{sec:QM_pipeline}. The computed molecule/reaction- and atom-level descriptors are both standardized, and the latter is additionally expanded into a vector representation through the application of a radial basis function (RBF) expansion (cf. Figure \ref{fig:model_architecture} for an overview of where the respective descriptors are introduced in the network). 

For the considered dataset of [3 + 2] cycloaddition activation and reaction energies, the benefit of QM-augmentation compared to the regular GNN baseline is limited; the improvement in accuracy as observed in 10-fold cross-validation experiments with random splits amounts to $\sim$5\% (mean absolute error of 2.64 vs. 2.77 kcal/mol for $G^{\ddagger}$ and 2.40 vs. 2.52 kcal/mol for $G_r$). As evident from the learning curves in Section \ref{sec:bayesian_opt}, and as is consistent with our prior observations for QM augmentation in general, the improvement in accuracy is somewhat more significant for small training set sizes and diminishes as the dataset grows. 
A more substantial improvement is observed when selective splits based on substituent identity are probed (improvements in the accuracy of $\sim$20-30\% in most cases; cf. Section \ref{sec:selective_sampling}), suggesting improved robustness and generalization ability of the model upon QM-augmentation -- which would be in line with the findings from previous work on this topic. \cite{stuyver2022quantum, beker2019prediction} 

Taking all these factors into account, we select the QM-augmented version of our ensembled GNN as the model with which to screen the large space of hypothetical bio-orthogonal reactions.

\subsection{Quantitative criteria to gauge bio-orthogonality potential}

Unambiguous quantitative thresholds to separate promising from unpromising candidates for bio-orthogonal click chemistry are difficult to determine due to the expected variability in the precise reaction conditions (temperature, concentrations, reaction environment, etc.) and the uncertainty of our models, as well as the intrinsic uncertainty related to the underlying DFT calculations. Nevertheless, one can get a broad sense of suitable criteria by benchmarking the computed reaction properties of known bio-orthogonal click reactions. 

In ref. \citenum{stuyver2022reaction}, we analyzed the bio-orthogonal click potential of the tried-and-true methyl and acyl azides (which we used as a reference point throughout this study). We observed that the vast majority of the reactions involving biologically inspired fragments tested exhibited activation energies well above 25 kcal/mol, with one reaction involving acyl azide reaching a reaction barrier of only 23.6 kcal/mol. The most reactive combinations of these dipoles and synthetic (strained) dipolarophiles -- mainly substituted cyclooctynes -- on the other hand involved activation energies of 18-22 kcal/mol.

Based on this data and the fact that our GNN reaches an out-of-sample accuracy of approximately 3 kcal/mol (\emph{vide infra}), we pruned all dipole-dipolarophile combinations involving dipoles for which our final model predicts activation energies below 22 kcal/mol for any biologically-inspired dipolarophile under the condition that the corresponding reaction is not endothermic by more than 3 kcal/mol. These combinations have virtually no chance of being selective, and hence they will likely never fulfill the qualitative criteria outlined in the Introduction. For the same reason of promoting selectivity, any dipolarophile that reaches a mean activation energy across all retained dipoles below 22 kcal/mol is discarded as well. 

Subsequently, we removed all the dipole-synthetic dipolarophile combinations for which the model predicts an activation energy above 25 kcal/mol, since the corresponding reactions are unlikely to be fast enough for practical applications. Additionally, we also impose that the corresponding reverse reaction should have a barrier exceeding 28 kcal/mol.

\begin{figure}
\centering
\includegraphics[scale=1]{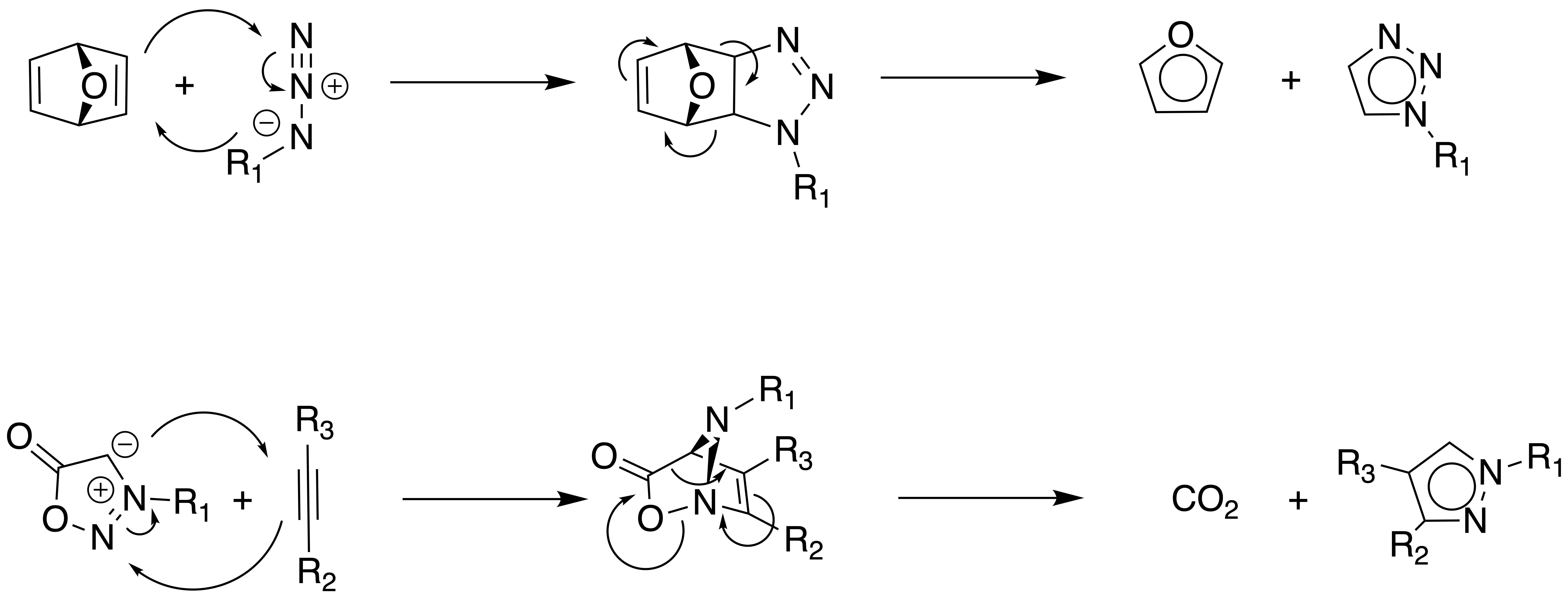}
\caption{Decomposition of an oxo-norbornadiene/azide addition product (top), giving rise to two delocalized -- and hence thermodynamically stabilized -- products, and decomposition of a sydnone/alkyne addition product, resulting in the formation of CO$_2$ and a (delocalized) triazole. Such pathways are not considered in the virtual screening pipeline used in this work.}
\label{fig:decomposition}
\end{figure}

A side note needs to be made here with regard to the final criterion mentioned above. It is known that the addition products of some of the dipole-dipolarophile combinations have the tendency to decompose along an alternative reaction mode, facilitated by a gain in aromaticity/delocalization energy \cite{sletten2009bioorthogonal, stuyver2020unifying}. More specifically, addition products involving oxo-norbornadiene can eject a furan, whereas cyclic dipoles can eject a CO$_2$ molecule -- particularly when reacting with an alkynyl-based dipolarophile (Figure \ref{fig:decomposition}). 

These pathways may affect the thermodynamic driving forces of the entire addition process. Probing these secondary reactions extends beyond the scope of this work, though one can reasonably expect the vast majority of the promising reactions involving these scaffolds to still pass this last filtering step due to the relatively permissive thermodynamic threshold value imposed (e.g., all the reactions involving oxo-norbornadiene and methyl/acyl azide computed in ref. \citenum{stuyver2022reaction} exhibited barriers for the reverse process of 40-50 kcal/mol). Nevertheless, one can expect some candidate click reactions that fail to fulfill the thermodynamic criterion listed above to irreversibly decompose in practice, resulting in an unwarranted rejection according to the adopted filtering procedure. Since the final model has been made available via GitHub, any examples in this gray zone can be subjected to further scrutiny in the future. 

For the dipole-dipolarophile combinations that passed all of the filtering steps outlined above, we quantified their bio-orthogonal click potential as the energy difference between the lowest barrier with a native (biological) dipolarophile and the barrier obtained for the considered synthetic (strained) dipolarophile. The bigger this energy difference, the more likely that that specific combination will correspond to a suitable bio-orthogonal click reaction. 

\subsection{Active learning procedure}

To establish confidence in the ability of our model to find appropriate hits -- and ensure the presence of both positive and negative examples in our training data -- we adopted a (semi-automated) active learning setup during the initial stages of our reaction discovery campaign. \cite{cohn1996active, reker2016multi} More specifically, we screened the entire search space several times. After a completed iteration, a sample of the most promising reactions was selected and the bio-orthogonal click potential of each reaction within the sample was validated through explicit DFT calculations, after which these newly acquired data points were added to the dataset and the model was retrained. 

Since the goal of the active learning loop consists of enriching the dataset with the most promising bio-orthogonal click reactions within the search space, we set more stringent selection criteria than above for this part of the study to minimize the number of false positives selected, i.e., to maximize the precision of the candidate selection procedure. The following paragraphs describe the protocol used for each iteration.

In a first step, the barriers for all the reactions between each of the dipoles and biologically relevant motifs (reaction SMILES generated through template matching with the help of RDChiral and RDKit) \cite{landrum2006rdkit, coley2019rdchiral} are predicted with the ML model trained in the previous iteration. Every dipole for which an activation energy lower than a threshold (set here at 26 kcal/mol) is predicted is then discarded -- under the condition that the corresponding reaction is not endothermic by more than 5 kcal/mol.

For the retained dipoles, the kinetics and thermodynamics of the reaction with each of the synthetic dipolarophiles is subsequently predicted. Every dipolarophile for which the mean activation energy across all reaction partners is lower than 26 kcal/mol is subsequently removed as this indicates nonspecific reactivity. 
From the remaining reactions, only those for which the activation energies in the forward direction are lower than 21 kcal/mol -- while being above 30 kcal/mol in the reverse direction -- are retained as the most promising candidates for bio-orthogonal click chemistry, and validated through the full DFT protocol.
With these selected thresholds, we are able to retain a manageable pool of candidate reactions, and a relatively high validation success rate is achieved across the iterations -- especially after the first round of dataset enriching (\emph{vide infra}).

Upon completion of the active learning loops, a final version of our ensembled GNN was trained based on the updated dataset enriched in promising bio-orthogonal click reactions. We then screened the search space one last time with relaxed criteria, enabling us to rank thousands of reactions based on the criteria listed above.

\section{Results and Discussion}

\subsection{Active learning improves the fidelity of the surrogate machine learning model}

Over the course of three rounds of iterative learning, our ML model gradually refines its predictions of energies in a subtle manner and increases its ability to distinguish promising dipole-dipolarophile combinations from unpromising ones. 

Already from the very first exploration round, the trained ML model is able to make accurate predictions across the entire search space, with an out-of-sample accuracy (evaluated on the reactions selected for click potential validation) on par with the accuracy obtained through cross-validation within the initial dataset. As the dataset gradually expands through successive iterations, the top-line accuracy only marginally improves (by 0.1-0.2 kcal/mol; cf. Tables \ref{tbl:performance_ensemble}-\ref{tbl:performance_ensemble_validation}). 
Despite the negligible effect of dataset expansion on top-line accuracy, the quality of the model significantly improves in a more subtle manner throughout the active learning procedure: the predictions become less conservative at the margins, causing many more reactions to pass the filtering steps (especially after the first iteration; see Table \ref{tbl:statistics_active_learning}).

\begin{table}
  \caption{Summary of some key statistics across the active learning iterations: the number of reactions passing all filtering steps, the number of distinct dipolarophiles and dipoles present in the resulting candidate pool and the mean absolute error (MAE) between computed and predicted $G^{\ddagger}$, obtained through 10-fold cross-validation on the full dataset in its current version, i.e., expanded by the unique validation reactions computed during preceding iterations.}
\label{tbl:statistics_active_learning}
  \begin{tabular}{ccccc}
    \hline
    iteration & \begin{tabular}{@{}c@{}} 
    \# of retained \\ reactions \end{tabular} & \begin{tabular}{@{}c@{}} 
    \# of distinct \\ dipolarophiles \end{tabular} & \begin{tabular}{@{}c@{}} 
    \# of distinct \\ dipoles \end{tabular} & \begin{tabular}{@{}c@{}} 
    MAE on $G^{\ddagger}$ for the \\ (expanded) dataset (kcal/mol) \end{tabular} \\
    \hline
    0 & 488 & 111 & 40 & 2.64 \\
    1 & 975 & 163 & 25 & 2.56 \\
    2 & 903 & 178 & 27 & 2.49 \\
    3 & 1123 & 221 & 30 & 2.47 \\
  \end{tabular}
\end{table}

Concomitant with the increase in retained reactions, the number of distinct dipolarophiles represented increases as well. The vast majority of dipolarophiles passing all the stringent filtering steps turn out to be based on a cyclooctyne scaffold (> 90\%); the oxo-norbornadiene and norbornene scaffolds are initially not represented. Attempts to increase the proportion of these alternative strained scaffolds by prioritizing their sampling during validation improves this situation somewhat for oxo-norbornadiene, but this scaffold still only features in around 1\% of the retained reactions in the last active learning iteration. Remarkably, even after explicit attempts to enrich the dataset with reactions involving norbornene-based dipolarophiles predicted to be the most reactive, no reaction with this scaffold passes all the active learning filtering steps. This is a somewhat surprising finding given the high strain energy of norbornene scaffolds \cite{khoury2004ring} and its use in other click reactions beyond the [3 + 2] cycloaddition reactions studied here \cite{devaraj2008tetrazine}.

\begin{figure}
\centering
\includegraphics[scale=0.8]{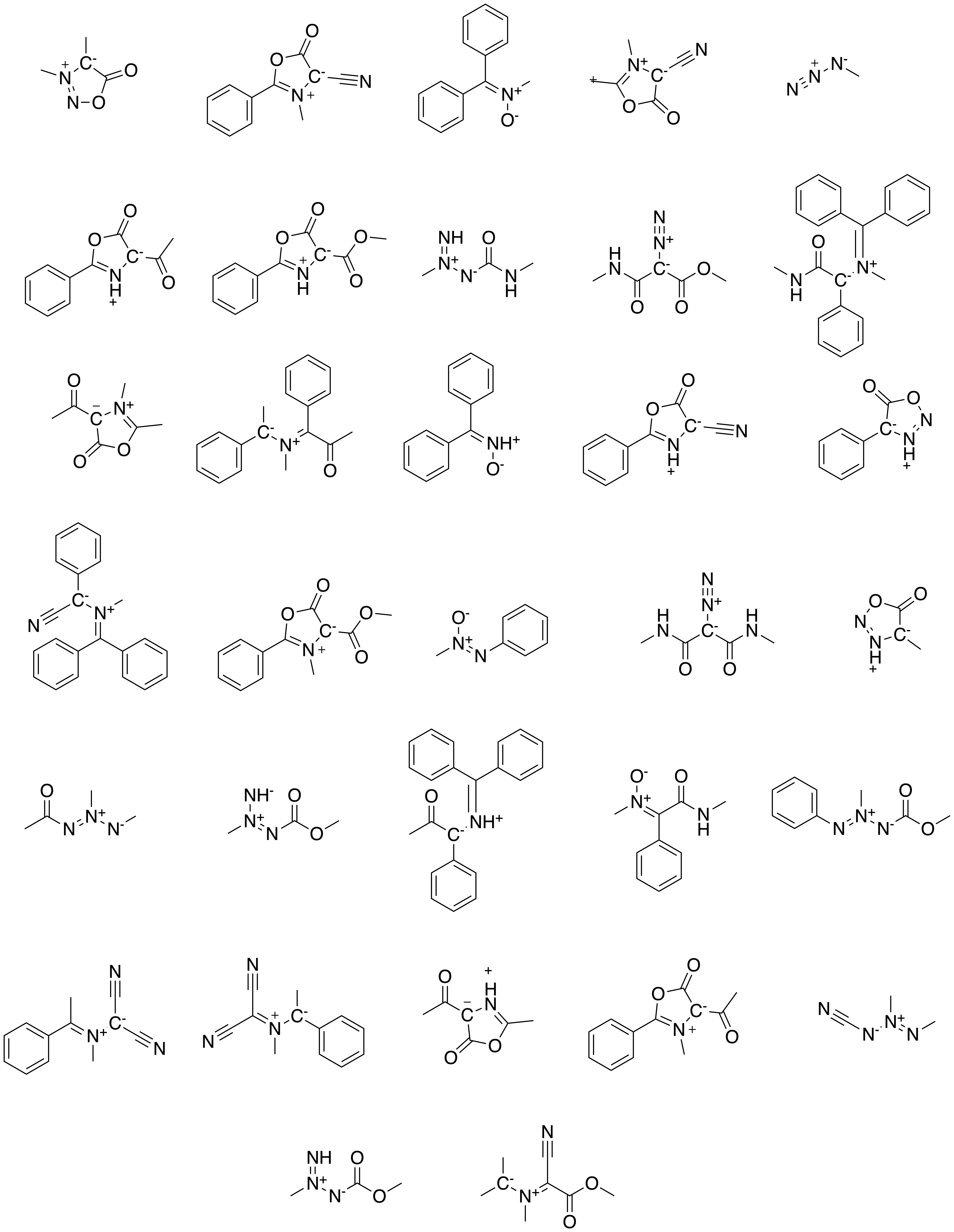}
\caption{Dipoles predicted by DFT to be relatively unreactive with each of the selected biologically inspired motifs. These species were selected as part of the validation effort during the active learning procedure.}
\label{fig:validated_dipoles}
\end{figure}

Next to an increase in the number of reactions retained by our models, we also observe a higher proportion of "true positives". While initially, only 25\% of the synthetic reactions sampled for validation exhibited more promising computed characteristics than the prototypical methyl and acyl azides discussed above, that rate rose to 66\% after the first iteration and reached approximately 40\% in the second iteration. In Figure \ref{fig:validated_dipoles}, an overview of the computationally validated dipoles can be found. Note that one can find examples of each dipole scaffold type in this Figure. 

\begin{figure}
\centering
\includegraphics[scale=0.5]{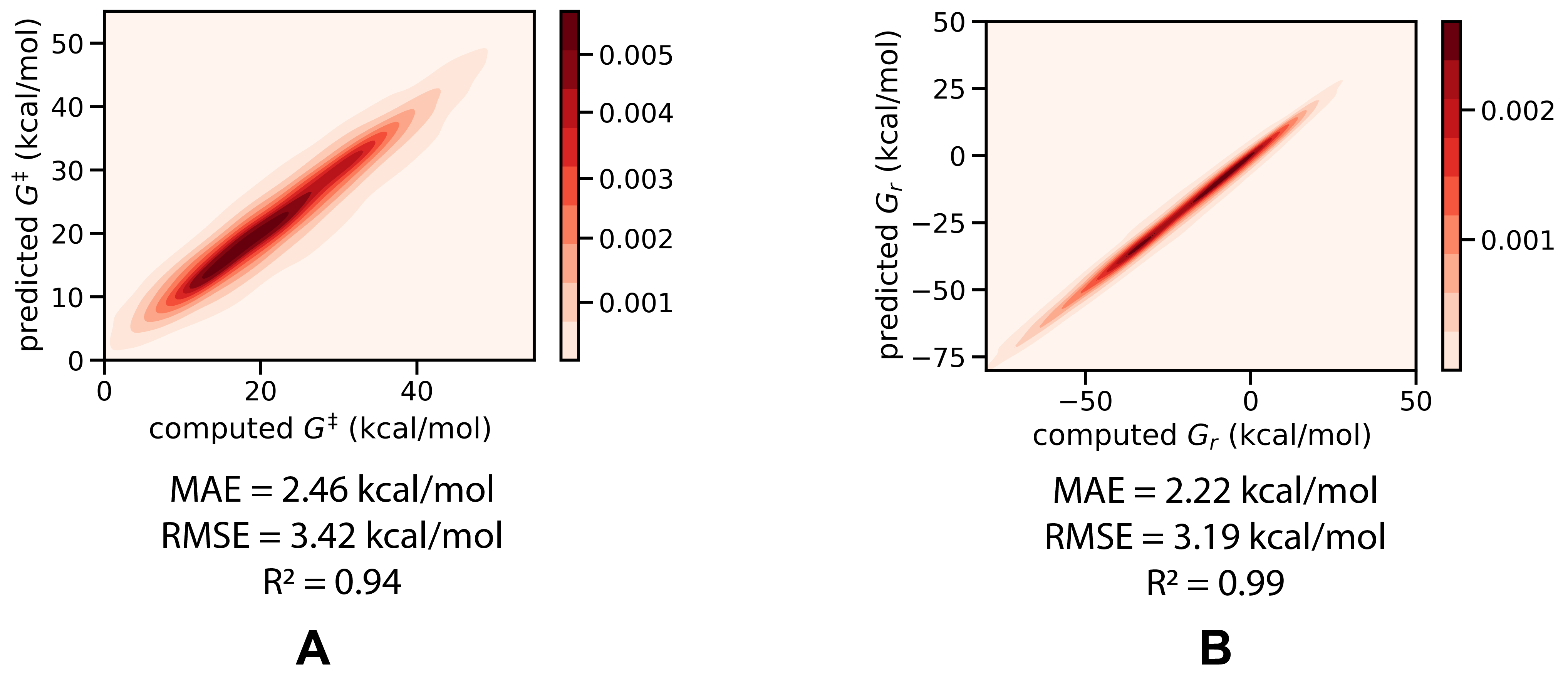}
\caption{Correlation plots between computed and predicted activation energies $G^{\ddagger}$ (A) and reaction energies $G_r$ (B) for the in-sample test sets considered during 10-fold cross-validation.}
\label{fig:correlation_plots}
\end{figure}

Correlation plots between the predicted and computed activation/reaction energies for the final model are presented in Figure \ref{fig:correlation_plots}. A more in-depth discussion of the individual iterations of the active learning procedure can be found in Section \ref{sec:active_learning_results}.

\subsection{Screening the search space for bio-orthogonal click potential with the refined model yields diverse candidates}

With the final instance of the QM-augmented ensemble model, we screened the entire search space for bio-orthogonal click potential according to the (more relaxed) criteria outlined in the Methodology Section. In total, 110,135 dipole-dipolarophile combinations within our search space pass all (relaxed) filtering steps, and 80,007 of these exhibit a positive bio-orthogonal potential, i.e., the activation energy for the fastest synthetic reaction involving this combination is predicted to be lower than the activation energy for any of the native reactions involving the same dipole. 

Among the retained combinations, we observe a wide variety in the dipole scaffold: 54\% of them involve a propargyl/allenyl-type dipole, 36\% an allyl-type dipole, and the remainder involves a cyclic dipole. Among the dipolarophiles, we observe that the vast majority of the retained dipolarophiles are cyclooctyne-based (81\%). The second most common dipolarophile scaffold is oxo-norbornadiene (10\%), followed by the non-strained dipolarophiles (8\%). While a handful of reactions involving a norbornene scaffold are retained now (554), their relative abundance is negligible (<1\%).

\begin{figure}
\centering
\includegraphics[scale=0.65]{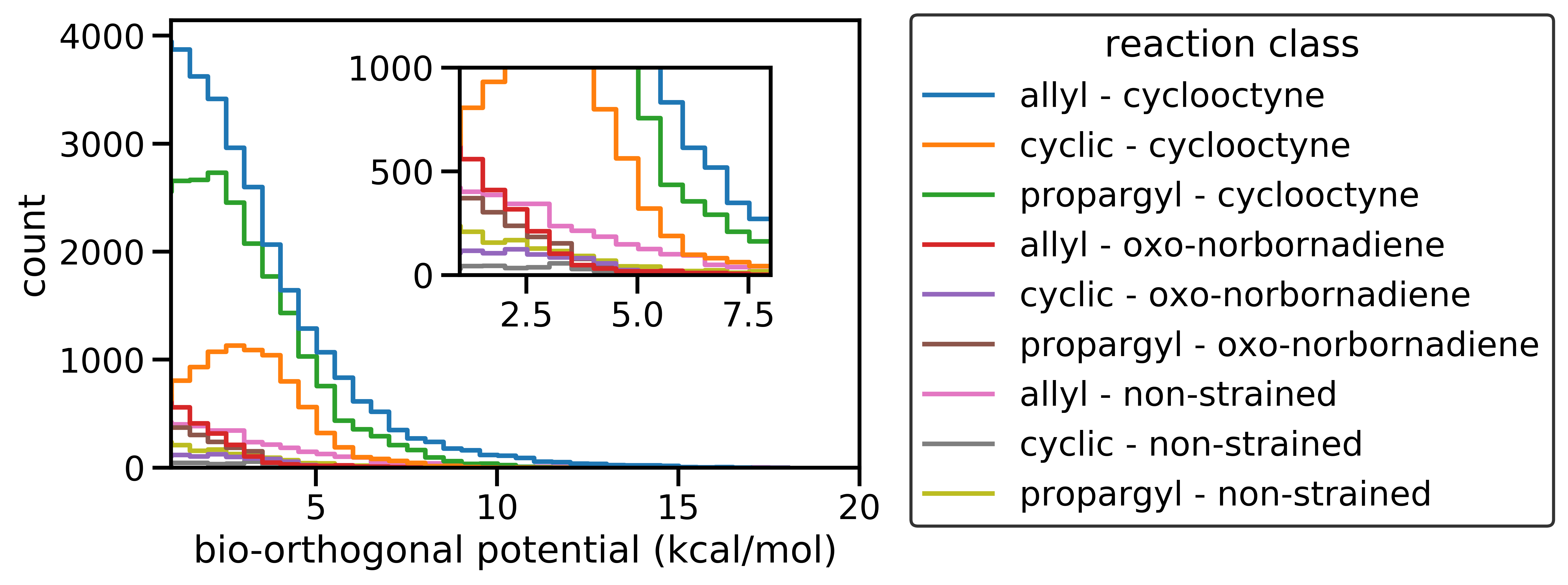}
\caption{Histogram depicting the distribution of the retained reactions with respect to their bio-orthogonal potential. Reaction classes are assigned based on dipole scaffold (allyl, propargyl or cyclic dipoles) and dipolarophile scaffold (cyclooctyne, oxo-norbornadiene and non-strained ethylenes/ethynes). Due to the negligible number of retained norbornene-based reactions, the classes involving the latter scaffold are not displayed in the histogram.} 
\label{fig:histogram}
\end{figure}

In Figure \ref{fig:histogram}, a graphical overview of the diversity found in the retained reactions is presented. Furthermore, we also provide the final trained model in the associated GitHub repository, together with a Jupyter Notebook containing an interactive visualization tool plotting the activation energy for the synthetic reactions as a function of the corresponding lowest native activation energy.

Some of the trends in the relative abundance of scaffolds are remarkable. First and foremost, while strain in the dipolarophile scaffold certainly helps increase the reactivity relative to non-strained native reactions, its presence is neither a sufficient nor a necessary condition for bio-orthogonal click potential: as discussed, norbornene, whose high strain energy is well-documented, \cite{khoury2004ring} failed to produce many promising candidate bio-orthogonal click reactions, whereas non-strained dipolarophiles based on ethylene/ethyne did produce a wide range of promising reactions.  

\begin{figure}
\centering
\includegraphics[scale=0.75]{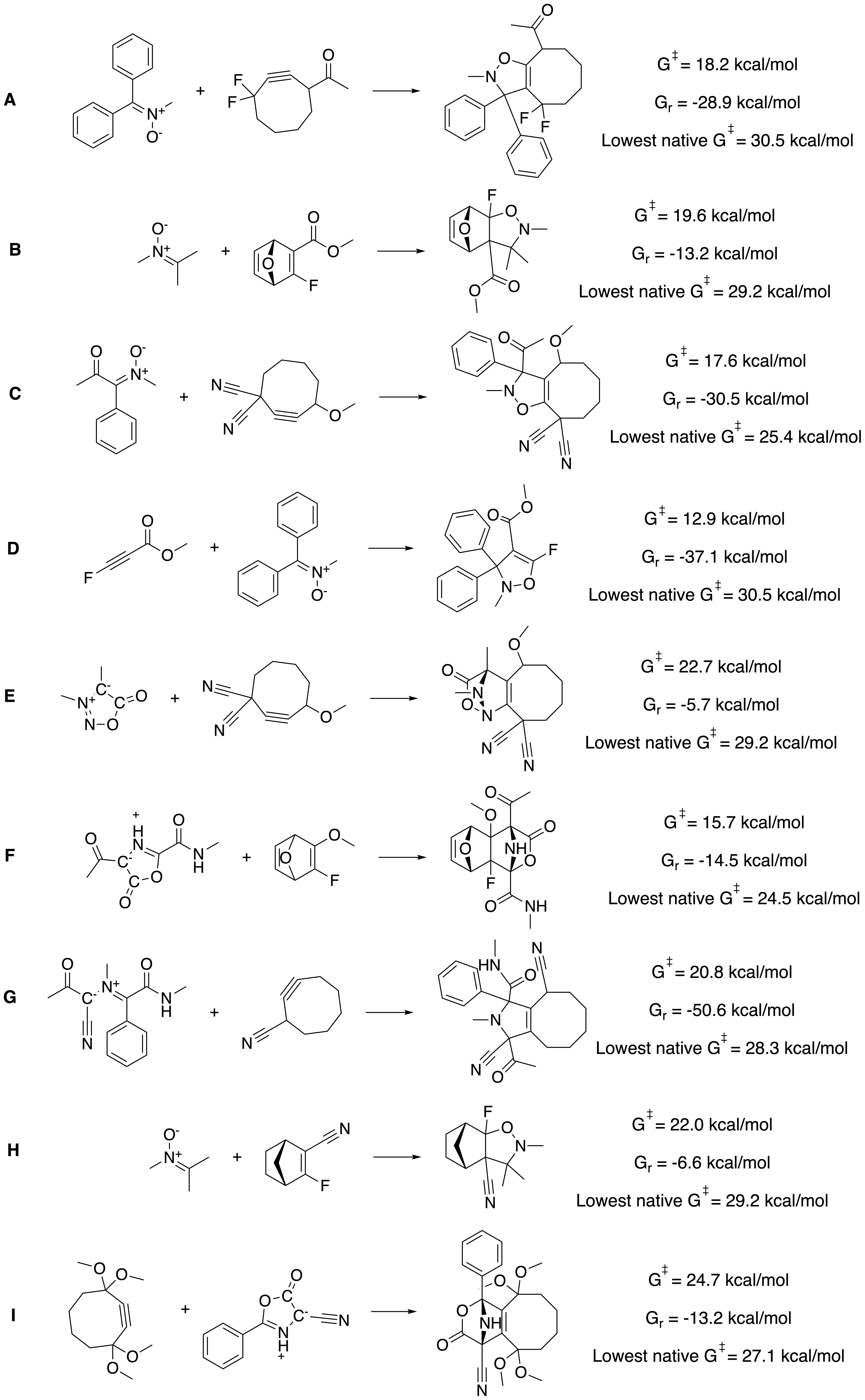}
\caption{Some reactions that are predicted to be suitable bio-orthogonal click reactions. (A-D) Promising nitrone-based reactions combining selectivity towards native reactions with excellent kinetic and thermodynamic properties. (E-F) Examples of promising reactions involving sydnone and m\"unchnone respectively. (G) A promising reaction based on a singly-substituted cyclooctyne dipolarophile. (H) The most promising reaction involving norbornene. (I) The most promising reaction involving only electron-donating groups.} 
\label{fig:promising_reactions}
\end{figure}

Focusing on the individual reaction level, one (sub)scaffold that immediately stands out are nitrone derivatives (Figure \ref{fig:promising_reactions}). While these species -- especially when substituted with methyl and/or phenyl groups -- are predicted to be relatively inert toward cycloaddition with the biologically inspired motifs tested here (lowest native $G^{\ddagger}$ typically above 25-30 kcal/mol), they reach excellent $G^{\ddagger}$ values with various types of synthetic dipolarophiles (many hundreds reach values significantly below 20 kcal/mol), suggesting that these reactions would transpire rapidly and selectively under physiological conditions (e.g., entries A-D and H in Figure \ref{fig:promising_reactions}). 

As mentioned in the introduction, nitrones have been considered experimentally before, so it is not entirely surprising that they show up in our analysis here. \cite{dommerholt2010readily, bilodeau2021bioorthogonal} Our results however suggest that they exhibit significantly more favorable characteristics for bio-orthogonal click chemistry than corresponding azide-based reactions in terms of inherent reactivity and that a broader range of dipolarophiles would be compatible with this species. Furthermore, some unexpected combinations appear feasible in principle for this scaffold: in fact, the nitrone-dipolarophile combinations exhibiting the most bio-orthogonal potential involve no strain release in the dipolarophile whatsoever and are based on an extremely compact substituted alkyne scaffold (entry D in Figure \ref{fig:promising_reactions}). It should be noted that some of these non-strained, alkyne-based, dipolarophiles -- particularly the fluorinated ones -- have been characterized in the literature as unstable and explosive in their own right, \cite{delavarenne1970heterosubstituierte} so that one cannot expect all the reactions of this type to be feasible in practice.

In addition to nitrones, our final screening also identifies plenty of reactions based on derivates of sydnone and m\"unchnone (as well as several other cyclic dipole scaffolds) as promising (e.g. entries E and F in Figure \ref{fig:promising_reactions}). Quite  a number of these reactions are however borderline cases, where our confidence about the actual bio-orthogonal potential is limited due to the uncertainty of the model. Interestingly, the reasons for this limited confidence are diametrically opposite for the two scaffolds. Whereas our predictions indicate that most sydnones ought to be selective, i.e., the lowest native $G^{\ddagger}$ tends to be predicted at 29-32 kcal/mol, the $G^{\ddagger}$ values they reach tend to be on the higher side -- typically in the range of 21-25 kcal/mol, casting doubt on whether all of these reactions will be fast enough for practical applications. For the m\"unchnones, on the other hand,  predicted $G^{\ddagger}$ values with the best synthetic dipolarophiles tend to be signficantly lower, often times reaching values around 15-18 kcal/mol, but the barriers with 
 the native dipolarophiles tend to be on the lower side (23-29 kcal/mol) as well, casting doubt on whether all of them will be selective. Substitution with a keto- or phenyl-group appears to temper the reactivity of m\"unchnone somewhat, causing the lowest native $G^{\ddagger}$ values to shift from the mid 20s to 26-29 kcal/mol, significantly increasing the likelihood that reactions based on these dipoles will exhibit an excellent selectivity.

We also find various reactions based on other allyl-based dipoles with promising kinetic and thermodynamic properties (e.g., entry G in Figure \ref{fig:promising_reactions}). As already discussed before, for the norbornene scaffold, the diversity in retained combinations is limited. Nevertheless, our model still finds several promising reactions based on this species (e.g., entry H).

A final note that needs to be made here is that the vast majority of the combinations retained involve electron-poor dipolarophiles and electron-rich dipoles, corresponding to so-called normal electron-demand Diels-Alder reactions (i.e., the main orbital interaction involves the dipole HOMO and dipolarophile LUMO). \cite{geittner1977kinetics} Even the best reactions with only electron-donating substituents on the dipolarophile, emerging from our screening campaign, do not have particularly promising properties (e.g. entry I in Figure \ref{fig:promising_reactions}). 

Our inability to find unequivocal inverse electron demand bio-orthogonal click reactions is likely the result of our search space definition: it appears that none of the 1,3-dipole scaffolds considered here can be rendered sufficiently electron-poor through substitution to rapidly react with electron-rich dipolarophiles. This also provides a plausible explanation for the limited representation of norbornene-based dipolarophiles in the set of retained combinations: the latter scaffold is found to be relatively electron-rich (its HOMO and LUMO respectively lie almost 0.7 and 0.5 eV higher in energy than the corresponding orbitals of the -- similarly localized -- cyclooctyne scaffold according to our calculations), so that matching its frontier orbitals with those from the considered dipoles likely becomes challenging. \cite{knall2013inverse} Including alternative scaffolds such as tetrazines -- which are known to react in a click manner with various types of electron-rich unsaturated compounds, e.g., various derivatives of norbornene \cite{blackman2008tetrazine,devaraj2008tetrazine} -- would likely remedy this situation, and could be the focus of future work. 

\section{Conclusions}
We presented here a computational workflow to discover new bio-orthogonal click reactions. We started by defining a broad search space including over 3,555 dipoles and almost 1,339 dipolarophiles, resulting in over 10,000,000 hypothetical dipolar cycloaddition reactions in total. Reaction profiles for a small subset of this search space were subsequently computed. Next, the resulting dataset was used to develop an ensembled GNN model, able to predict activation and reaction energies across the entire space. This GNN model was refined through 3 active learning iterations after which the search space was screened one final time. 

The resulting candidate pool of over 110,000 promising bio-orthogonal click reactions contains a rich structural diversity. 54\% of the reactions involve a propargyl/allenyl-type dipole, 36\% an allyl-type dipole and 10\% a cyclic dipole. Among the dipolarophiles, 81\% are based on the cycloocytne scaffold, 10\% involve the oxo-norbornadiene scaffold, 8\% are based on non-strained dipolarophiles and <1\% on the norbornene scaffold. A particular (sub)scaffold that stands out is nitrone. Various derivatives of this species are predicted to react rapidly with a broad range of dipoles, while being relatively unreactive with potentially competing native, i.e., biological, dipolarophiles.

The vast majority of the reactions discovered here involve electron-poor dipolarophiles and electron-rich dipoles. The lack of promising reactions exhibiting an inverse electron demand is likely due to our inability to render the considered dipoles sufficiently electron-poor through substitution to
rapidly react with electron-rich dipolarophiles. This also provides a plausible explanation for the limited representation of norbornene-based dipolarophiles in the set of retained combinations since the latter scaffold is found to be relatively electron-rich in our calculations.

While experimental evidence corroborates the bio-orthogonal potential of some of the identified candidate dipole-dipolarophile combinations (e.g., the nitrones), \cite{dommerholt2010readily, bilodeau2021bioorthogonal} one cannot expect all the reactions proposed here to work in practice. Experimental validation is important, and additional considerations for practicality in bio-orthogonal click chemistry that are beyond the scope of this computational pipeline need to be considered as well. Nevertheless, our work presented here demonstrates that machine learning can help us to leverage first principle calculations to explore massive combinatorial design spaces efficiently and provide promising starting points for discovery campaigns.

The approach presented above can in principle be expanded to any reaction type, as long as the ligation mechanism is unambiguous, consistent across the search space, and can be modeled accurately and economically; reaction profile computations were the bottleneck in this project, but these may be accelerated with neural potentials in the future. \cite{smith2019approaching, gokcan2022learning} 

\begin{acknowledgement}

TS and CWC thank the Machine Learning for Pharmaceutical Discovery and Synthesis Consortium for financial support. The authors acknowledge the MIT SuperCloud and Lincoln Laboratory Supercomputing Center for providing HPC resources that have contributed to the research results reported within this paper.

\end{acknowledgement}

\begin{suppinfo}

In-depth technical description of the GNN model architectures, hyperparameter search and model performance, comparison between the performance of the designed neural network architecture and alternative model types, motivation for the QM-descriptor selection and theoretical background, description of the QM-descriptor pipeline, comparison between the performance of the QM-augmented and regular/baseline GNNs for selective sampling experiments, in-sample model performance across active learning iterations, validation performance across active learning iterations, in-depth discussion of the active learning results

The code described in the previous section is freely available on GitHub under the MIT license (\url{https://github.com/coleygroup/bio_orthogonal_click_reactions}). Further details on how to use it is provided in the associated README.md file.

\end{suppinfo}


\putbib[biblio_ms]
\end{bibunit}

\clearpage
\newpage

\begin{bibunit}[unsrt]

\setcounter{section}{0}
\setcounter{equation}{0}
\setcounter{figure}{0}
\setcounter{table}{0}
\section*{Supplementary Information}
\normalsize
\setcounter{section}{0}
\renewcommand\thesection{S\arabic{section}}
\def\thefigure{S\arabic{figure}}
\def\thetable{S\arabic{table}}
\def\theequation{S\arabic{equation}}

\section{In-depth technical description of the GNN model architectures}\label{sec:architecture}

\subsection{The regular GNN model}

The first step that is taken when a new data point is parsed by the model is a conversion of the SMILES representations of reactants and products to graph-based/structural ones. More specifically, each atom $v$ is initialized with a feature vector $f_v$ indicating its atomic number, degree of connectivity, explicit and implicit valence, and aromaticity. Each bond (u, v) is associated with a feature vector $f_{uv}$ indicating its bond type and ring status. Subsequently, the structural representations are individually fed to a common Weisfeiler-Lehman network (WLN), which constructs a convolutional embedding of the atoms by iteratively updating their respective structural representation with information from adjacent atoms. In iteration \emph{t}, the atom representation at atom $v \epsilon G$ with neighbor nodes $N(v)$ is updated from $\mathbf{f}_v^{t-1}$ to $\mathbf{f}_v^t$ by first constructing an update vector $\mathbf{l}_v^t$. $\mathbf{l}_v^t$ is obtained by passing the concatenation of neighboring atom features, $\mathbf{f}_u^{t-1}$, and the corresponding bond features $\mathbf{f}_{uv}$, through a feedforward layer, followed by summation:

\begin{equation}
\mathbf{l}_v^t = \sum_{u \epsilon N(v)} \text{ReLU}(\mathbf{U}_2[\mathbf{f}_u^{t-1}; \mathbf{f}_{uv}]) 
\end{equation}

where RELU is the rectified linear unit and $\mathbf{U}_2$ is the learned matrix associated with the feedforward layer. Next, the resulting vectors are concatenated to $\mathbf{f}_v^{t-1}$ and passed through a second layer (with $\mathbf{U}_1$ the associated learned matrix):

\begin{equation}
\mathbf{f}_v^t = \text{ReLU}(\mathbf{U}_1([\mathbf{f}_v^{t-1}; \mathbf{l}_v^t]))
\end{equation}

After \emph{t} iterations (based on our hyperparameter search, 2 consecutive layers were selected; \emph{vide infra}), the final local atomic representations, $\mathbf{c}_v$ are computed as follows,
\begin{equation}
\mathbf{c}_v = \mathbf{V}_1 \mathbf{f}_v^{t} \odot \sum_{u \epsilon N(v)} \mathbf{W}_1 \mathbf{f}_v^{t-1} \odot \mathbf{W}_2 \mathbf{f}_{uv}
\end{equation}
where $\odot$ denotes the Hadamard product and $\mathbf{V}_1$, $\mathbf{W}_1$ and $\mathbf{W}_2$ are learned matrices.

The atomic embeddings $\mathbf{c}_v$ only encode local structural patterns, namely atoms and bonds accessible within $t$ steps from atom $v$. To capture distant information (e.g., information between disconnected atoms), the resulting embeddings $\mathbf{c}_v$ are then passed through an attention layer, which calculates the so-called "attention score" of atom $v$ on atom $z$. Attention scores for atom pair ($v$,$z$), $\mathbf{\alpha}_{vz}$, are calculated by this layer as follows,
\begin{equation}
\mathbf{\alpha}_{vz} = \sigma(\mathbf{Q}_1 \text{ReLU}(\mathbf{P}_1(\mathbf{c}_v + \mathbf{c}_z) + \mathbf{P}_2 \mathbf{b}_{vz})
\end{equation}
where $\sigma$ indicates a sigmoid activation function, $\mathbf{P}_1$, $\mathbf{P}_2$ and $\mathbf{Q}_1$ are learned matrices and $\mathbf{b}_{vz}$ are binary features describing the atom-atom relationships (e.g., the bond type between the atoms, if any, and whether they belong to the same molecule). The "global" atom representation $\mathbf{\tilde{c}}_v$ is defined as the weighted sum of all reactant atom feature vectors,
\begin{equation}
\mathbf{\tilde{c}}_v = \sum_z \mathbf{\alpha}_{vz} \mathbf{c}_z
\end{equation}
The final atomic representation for atom $v$, $\mathbf{\hat{c}}_v$, is then constructed as follows,
\begin{equation}
\mathbf{\hat{c}}_v = \text{ReLU}(\mathbf{M}(\mathbf{\tilde{c}}_v + \mathbf{c}_v))
\end{equation}
where \textbf{M} is again a learned matrix. This final atomic representation is then passed through a sum-pooling layer, which sums over the atomic representations of the atoms which undergo a change in bonding situation throughout the reaction, i.e., the atoms part of the reacting core ($RC$), resulting in the reaction representation. From this common reaction representation, the network splits up into two branches, respectively corresponding to the activation and reaction energy. Both branches consist of a feedforward layer:
\begin{equation}
\mathbf{s}_1 = \mathbf{O}_1(\sum_{u \epsilon RC} \mathbf{\hat{c}}_v)
\end{equation}

\begin{equation}
\mathbf{s}_2 = \mathbf{O}_2(\sum_{u \epsilon RC} \mathbf{\hat{c}}_v)
\end{equation}

with $\mathbf{O}_1$ and $\mathbf{O}_2$ learned matrices and $\mathbf{s}_1$ and $\mathbf{s}_2$ the (hidden) reaction vectors. Both of the resulting reaction vectors are finally projected to individual read-out scores, $s_{final,1}$ and $s_{final,2}$, with the help of learned matrices $\mathbf{O}_3$ and $\mathbf{O}_4$:
\begin{equation}
s_{final,1} = \mathbf{O}_3(\mathbf{s}_1)
\end{equation}

\begin{equation}
s_{final,2} = \mathbf{O}_4(\mathbf{s}_2)
\end{equation}

The loss function used for the model is the sum of mean squared errors for both individual targets.

\subsection{The ml-QM-GNN model}

As mentioned in the main text, the QM-augmented GNN model follows a similar architecture as the baseline GNN model, with the main difference being the introduction of QM-descriptors at specific points in the network. 

As indicated in the main text, the descriptors fed to the network can be subdivided in atom- and reaction-level descriptors. The atom-level descriptors consist of partial charges, and electrophilic/nucleophilic Fukui functions, as well as NMR shielding constants, \cite{wolinski1990efficient} whereas the reaction-level descriptors correspond to the respective promotion gaps derived from Valence Bond theory \cite{shaik2007chemist} (\emph{vide infra}).

All descriptors are normalized with the help of standard normalization. For the NMR shielding constants, the scaling was performed on an element-by-element basis due to the dramatic differences in shielding constant magnitudes depending on the atom-type, which completely dwarf the typical intra-element variation. 

Atomic-level descriptors are turned into vector representations with the help of a radial basis function (RBF) expansion. For continuous atom-centered descriptor, $a_u$, the vector representation $e_u$ is expressed as follows,
\begin{equation}
e_{u} = [\exp{(-\frac{(a_{u} - (\mu + \delta k))^2}{\delta})}]_{k\epsilon[0,1,2,...,n]}
\end{equation}
For each descriptor, the $\mu$, $\delta$ and $n$ parameters were set to -6.0, 0.6, and 20 respectively.  

As indicated in the main text, the vector representations of the (scaled) atom-level descriptors are first concatenated,

\begin{equation}
\hat{c}_v^{QM} = [a_u^q; a_u^{f+}; a_u^{f-}; a_u^{sc}]
\end{equation}

where $a_u^q$ is the partial charge RBF expanded vector, $a_u^{f+}$ and $a_u^{f-}$ are respectively the nucleophilic and electrophilic Fukui function RBF expanded vectors and $a_u^{sc}$ is the shielding constant RBF expanded vector. Subsequently, these QM representations are combined with the learned atomic representations emerging from the WL branch, i.e.,

\begin{equation}
\hat{c}_v^{QM-GNN} = [(1-\alpha_1)*\hat{c}_v; \alpha_1*\hat{c}_v^{QM}]
\end{equation}

where $\alpha_1$ is a fixed hyperparameter that controls the admixture of both representations. The resulting concatenated representation is then passed through the attention mechanism and an additional dense layer, followed by sum pooling to yield the reaction-level representation, $\hat{h}^{GNN}$. Next, the reaction-level descriptors are concatenated, 

\begin{equation}
\hat{h}^{QM} = [G; G^*; G^{**}]
\end{equation}

where G, $G*$, $G**$ correspond to the respective promotion gaps. The resulting vector is then concatenated to the reaction-level representation,

\begin{equation}
\hat{h}^{QM-GNN} = [(1-\alpha_2)*\hat{h}^{GNN}; \alpha_2*\hat{h}^{QM}]
\end{equation}

where $\alpha_1$ is a fixed hyperparameter that controls the admixture of both representations. This final representation is then passed through a final network of dense layers, after which a final projection to the two read-out scores, corresponding to activation and reaction energy, is performed, in a similar manner as in the regular GNN.

\section{Hyperparameter search and model performance}\label{sec:bayesian_opt}

In order to select a reasonable set of hyperparameters for our QM-augmented GNN, a minimal Bayesian optimization search for the latter was performed. 80\% of the data set was sampled and the average root-mean-square error on the activation energy emerging from 4-fold cross-validation on this sampled subset was set as the objective to minimize. A 7-dimensional search space was defined (cf. Table \ref{tbl:bayesian} for an overview of the hyperparameters selected for optimization) and 64 iterations were performed. The optimal parameters are shown in the final column of Table \ref{tbl:bayesian}.

\begin{table}
  \caption{Definition of the search space and the optimal parameter values emerging from the Bayesian optimization.}
  \label{tbl:bayesian}
  \begin{tabular}{ccccc}
    \hline
    hyperparameter & min & max & interval & optimal \\
    \hline
    depth WLN & 2 & 6 & 1 & 2 \\
    weight factor atom vectors & 0.1 & 1 & 0.1 & 0.5 \\
    weight factor reaction vector & 0.1 & 1 & 0.1 & 0.3 \\
    initial learning rate & $e^{-10}$ & $e^{-5}$ & \emph{log uniform} & 0.00165 \\
    learning rate ratio & 0.9 & 0.99 & 0.01 & 0.93 \\
    depth FFNN & 1 & 4 & 1 & 1 \\
    hidden size multiplier & 0 & 20 & 10 & 0 \\
    \hline
  \end{tabular}
\end{table}

In Table \ref{tbl:performance} the accuracy of our QM-augmented GNN with the determined optimal settings is presented. The regular GNN baseline is also included in this table. 

\begin{table}
  \caption{Comparison between the performance of the QM-augmented GNN and its regular analog, in terms of mean absolute error (MAE) and root mean square error (RMSE).}
  \label{tbl:performance}
  \begin{tabular}{cccc}
    \hline
    model type & target & MAE (kcal/mol) & RMSE (kcal/mol) \\
    \hline
    QM-augmented GNN ensemble & \begin{tabular}{@{}c@{}} 
    $G^{\ddagger}$ \\ $G_r$ \end{tabular} & \begin{tabular}{@{}c@{}} 2.96 \\ 2.89 \end{tabular} & 
    \begin{tabular}{@{}c@{}} 4.03 \\ 4.22 \end{tabular} \\ 
    \hline
    regular GNN ensemble & \begin{tabular}{@{}c@{}} 
    $G^{\ddagger}$ \\ $G_r$ \end{tabular} & \begin{tabular}{@{}c@{}} 3.09 \\ 3.09 \end{tabular} & 
    \begin{tabular}{@{}c@{}} 4.17 \\ 4.48 \end{tabular} \\ 
    \hline
  \end{tabular}
\end{table}

Additionally, learning curves for both targets have been determined for both the QM-augmented GNN and regular GNN baseline. Across the entire spectrum of training set sizes considered, one observes that the QM-augmented GNN slightly -- yet consistently -- outperforms the baseline. The improvement is the largest for the smaller training set sizes. 

\begin{figure}
\centering
\includegraphics[scale=0.8]{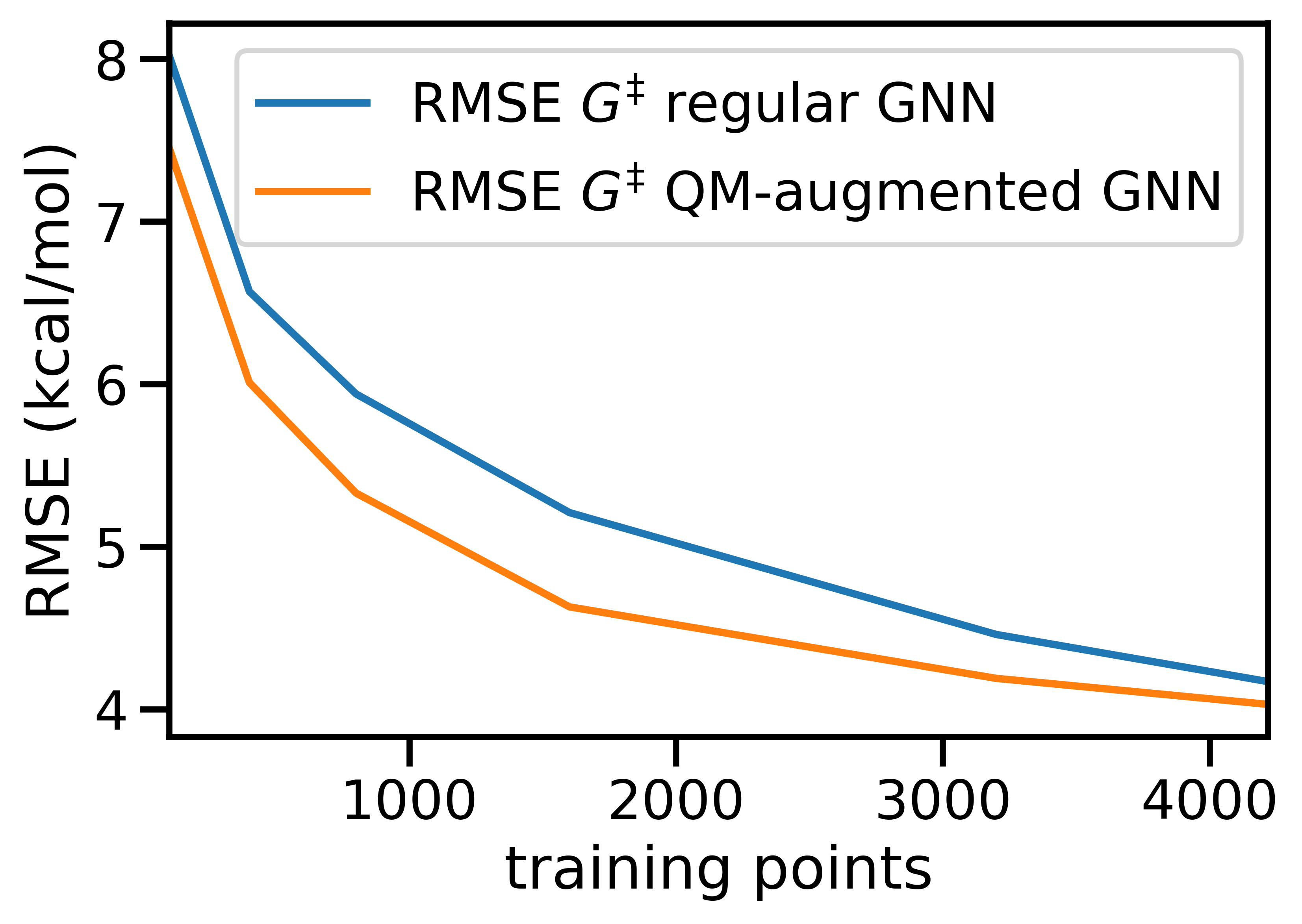}
\caption{Learning curves for the $G^{\ddagger}$ target for the QM-augmented and regular GNNs respectively.}
\label{fig:learning_curves_Gact}
\end{figure}

\begin{figure}
\centering
\includegraphics[scale=0.8]{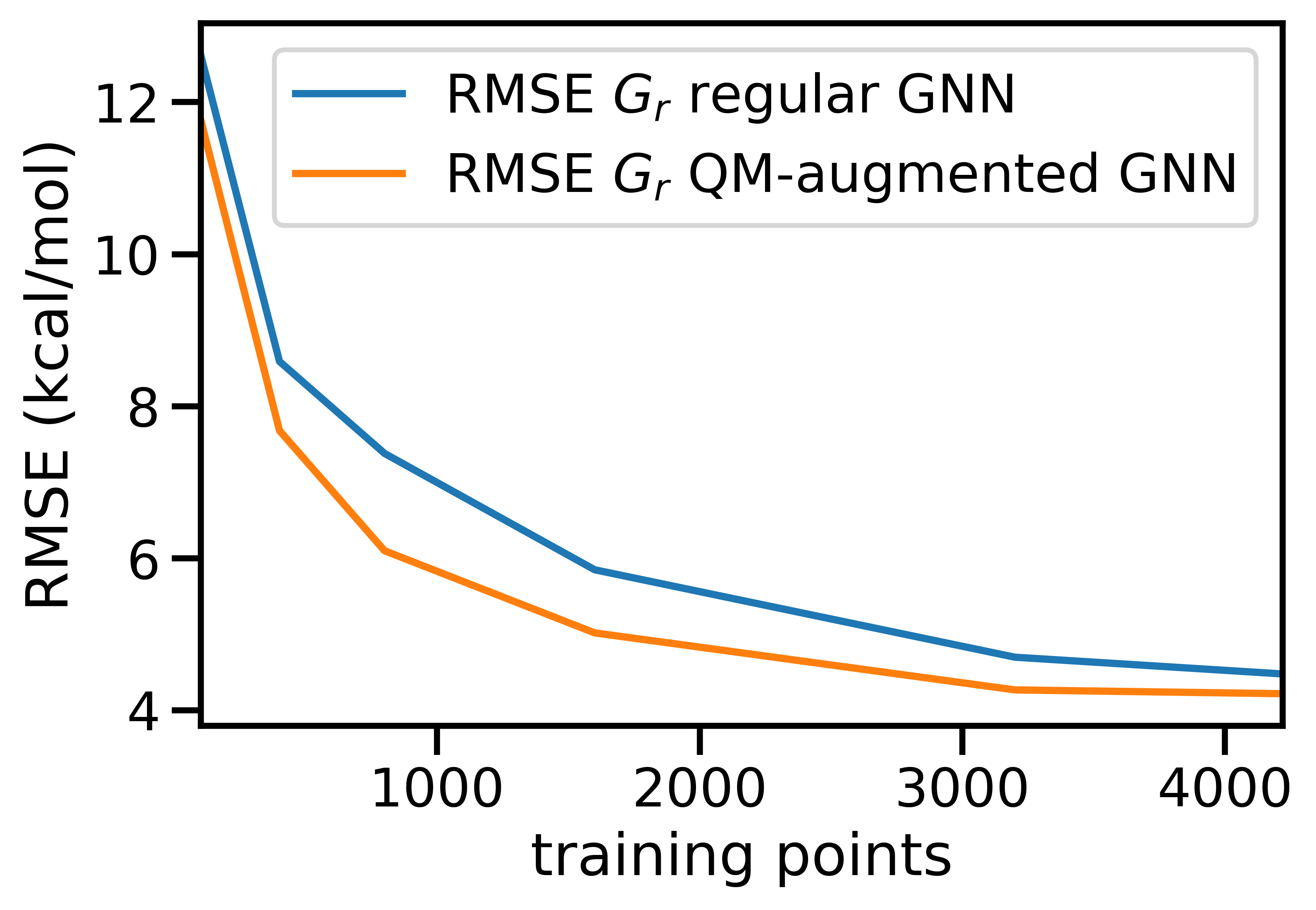}
\caption{Learning curves for the $G_r$ target for the QM-augmented and regular GNNs respectively.}
\label{fig:learning_curves_Gr}
\end{figure}

Finally, we also constructed an ensemble model of 10 GNNs with the selected hyperparameters and performed a 10-fold cross-validation. The obtained accuracy is summarized in Table \ref{tbl:performance_ensemble1}. This is the final model architecture that has been used throughout the remainder of the study.

\begin{table}
  \caption{Final performance, in terms of mean absolute error (MAE) and root mean square error (RMSE), of the ensemble model constructed from 10 QM-augmented GNNs.}
  \label{tbl:performance_ensemble1}
  \begin{tabular}{cccc}
    \hline
    model type & target & MAE (kcal/mol) & RMSE (kcal/mol) \\
    \hline
    QM-augmented GNN ensemble & \begin{tabular}{@{}c@{}} 
    $G^{\ddagger}$ \\ $G_r$ \end{tabular} & \begin{tabular}{@{}c@{}} 2.64 \\ 2.40 \end{tabular} & 
    \begin{tabular}{@{}c@{}} 3.63 \\ 3.64 \end{tabular} \\ 
    \hline
    regular GNN ensemble & \begin{tabular}{@{}c@{}} 
    $G^{\ddagger}$ \\ $G_r$ \end{tabular} & \begin{tabular}{@{}c@{}} 2.76 \\ 2.52 \end{tabular} & 
    \begin{tabular}{@{}c@{}} 3.77 \\ 3.80 \end{tabular} \\ 
    \hline
  \end{tabular}
\end{table}

\section{Comparison between the performance of the designed neural network architecture and alternative model types}\label{sec:alternative_algorithms}

Several alternative model architectures were tested. To facilitate comparison with the GNN results, the accuracy on the activation energy task $G^{\ddagger}$ was determined in 10-fold cross-validation on the same splits as used for the GNN (Table \ref{tbl:performance_alt_models}). Whenever applicable, hyperparameters were selected through a Bayesian Optimization-based (BO) search. During BO, a test set of 20\% of the data was held out, after which 4-fold cross-validation on the remaining data was performed. The average root mean square error across the different folds was selected as the metric to optimize.

The first model architecture tested is simple linear regression based on descriptor values. For the atom-level descriptors, each of the 4 descriptors on the 5 active sites, i.e., the atoms which undergo a change in bonding throughout the reaction, were selected to this end. Additionally, the 3 reaction-level descriptors were also included, yielding 23 input features in total. 

A component-based k-nearest neighbors model was also set up. In this model, Euclidian distances between Morgan fingerprint-based representations (radius = 2; nbits = 2048) of dipole, dipolarophile and product respectively for each test-train reaction combination are determined, and a weighted average of each distance is taken. The prefactors for the individual distances were defined with the help of 2 hyperparameters ($\lambda$ and $\mu$): $\lambda$, $(1 - \lambda) * \mu$, $(1 - \lambda) * (1 - \mu)$ respectively. Additionally, the number of neighbors ($k$) was set as a hyperparameter as well. The best parameters across 64 iterations of Bayesian Optimization were $\lambda = 0.4$, $\mu = 0.1$ and $k = 4$.

The next model architecture tested is a random forest. Both descriptors and Morgan fingerprints (radius = 2; nbits = 2048) were considered as features. The hyperparameters, maximal fraction of features ($max_{features}$), minimal number of samples per leaf ($min_{samples-leaf}$), and the number of estimators ($n_{estimators}$), were determined with the help of Bayesian Optimization. For the descriptor-based featurization, optimal performance across 32 iterations was achieved for $max_{features} = 0.5$, $min_{samples-leaf}$ = 2 and $n_{estimators} = 263$. For the fingerprint-based featurization, optimal performance was achieved for $max_{features} = 0.2$, $min_{samples-leaf}$ = 2 and $n_{estimators} = 245$.

A final model architecture considered was XGBoost. Here as well, both descriptors and Morgan fingerprints (radius = 2; nbits = 2048) were considered as features. The hyperparameters in this case are $\gamma$, learning rate ($lr$), maximal depth of a tree ($max_{depth}$), minimum sum of instance weight needed in a child ($min_{child-weight}$) and the number of estimators ($n_{estimators}$). After 128 iterations of Bayesian optimization, optimal hyperparameters were found to be $\gamma$ = 2.0, $lr$ = 0.1, $max_{depth}$ = 9, $min_{child-weight}$ = 6 and $n_{estimators}$ = 800 for the descriptor-based featurization. For the fingerprint-based featurization, optimal hyperparameters were found to be $\gamma$ = 2.0, $lr$ = 0.2, $max_{depth}$ = 9, $min_{child-weight}$ = 8 and $n_{estimators}$ = 500.

\begin{table}
  \caption{Performance on $G^{\ddagger}$, in terms of mean absolute error (MAE) and root mean square error (RMSE), of the alternative model architectures tested.}
  \label{tbl:performance_alt_models}
  \begin{tabular}{ccc}
    \hline
    model & MAE (kcal/mol) & RMSE (kcal/mol) \\
    \hline
    linear regression (descriptors)  & 5.94 & 7.73 \\
    k-nearest neighbors (fingerprints) & 4.33 & 5.74 \\
    random forest (descriptors) & 3.73 & 5.06 \\
    random forest (difference fingerprint) & 3.50 & 4.76 \\
    XGBoost (descriptors) & 3.78 & 5.05 \\
    XGBoost (difference fingerprints) & 3.84 & 5.07 \\
    \hline
  \end{tabular}
\end{table}

Table \ref{tbl:performance_alt_models} demonstrates that all our alternative model architecture manage to learn trends within the dataset: they beat the simplest baseline of all, i.e., always predicting the mean value, by several kcal/mol (standard deviation for the whole dataset amounts to 9.8 kcal/mol, cf. Figure 1 in the main text). Nevertheless, none of the architectures tested comes even close to the performance of the GNN/QM-augmented GNN (the RMSE of the best alternative model, random forest based on difference fingerprints, is still 0.5-0.75 kcal/mol higher), and thus, we retained the latter architecture as the base model for our ensemble.

\section{Motivation for the QM-descriptor selection and theoretical background}\label{sec:QM_descriptor_motivation}

With the atom-level descriptors, we aimed to provide the GNN with rich information about the main interaction types that govern chemical reactivity: electrostatic, i.e., hard-hard, and orbital, i.e., soft-soft, interactions. \cite{pearson1963hard, stuyver2020unifying}

To describe the electrostatics of the reactants, we selected partial charges as well as NMR shielding constants. The partial charges were computed according to the Hirshfeld partitioning scheme since this scheme has only a small dependency on basis sets, and Hirshfeld charges are able to reproduce the electrostatic potential, which are important to reactivity and molecular properties. \cite{wiberg1993comparison, fuentealba2000condensed}

NMR shielding constants provide information about local deformations of the electron cloud around individual nuclei and these values indirectly probe polarizability. They were calculated with the Gauge-Independent Atomic Orbital (GAIO) method. \cite{wolinski1990efficient, helgaker1999ab}

To cover the orbital interactions between the reactants as they react, we selected the nucleophilic and electrophilic Fukui functions. \cite{geerlings2003conceptual} Taking the finite-difference approximation, the nucleophilic Fukui function, $f_i^-$ can be expressed as follows,

\begin{equation}
f^-_i = q_i(N-1) - q_i(N)
\end{equation}

and the electrophilic Fukui function, $f_i^+$, can be expressed as follows,

\begin{equation}
f^+_i = q_i(N) - q_i(N+1)
\end{equation}

where we again computed the charges with the help of the Hirshfeld partitioning scheme. \cite{hirshfeld1977bonded}

For the molecule-level descriptors, we selected the various promotion gaps which can be defined within a Valence Bond perspective. \cite{shaik2007chemist} The heights of these promotion gaps have previously been demonstrated to be intimately related to the magnitude of reaction barriers. \cite{shaik2007chemist, stuyver2021promotion}

\begin{figure}
\centering
\includegraphics[scale=1]{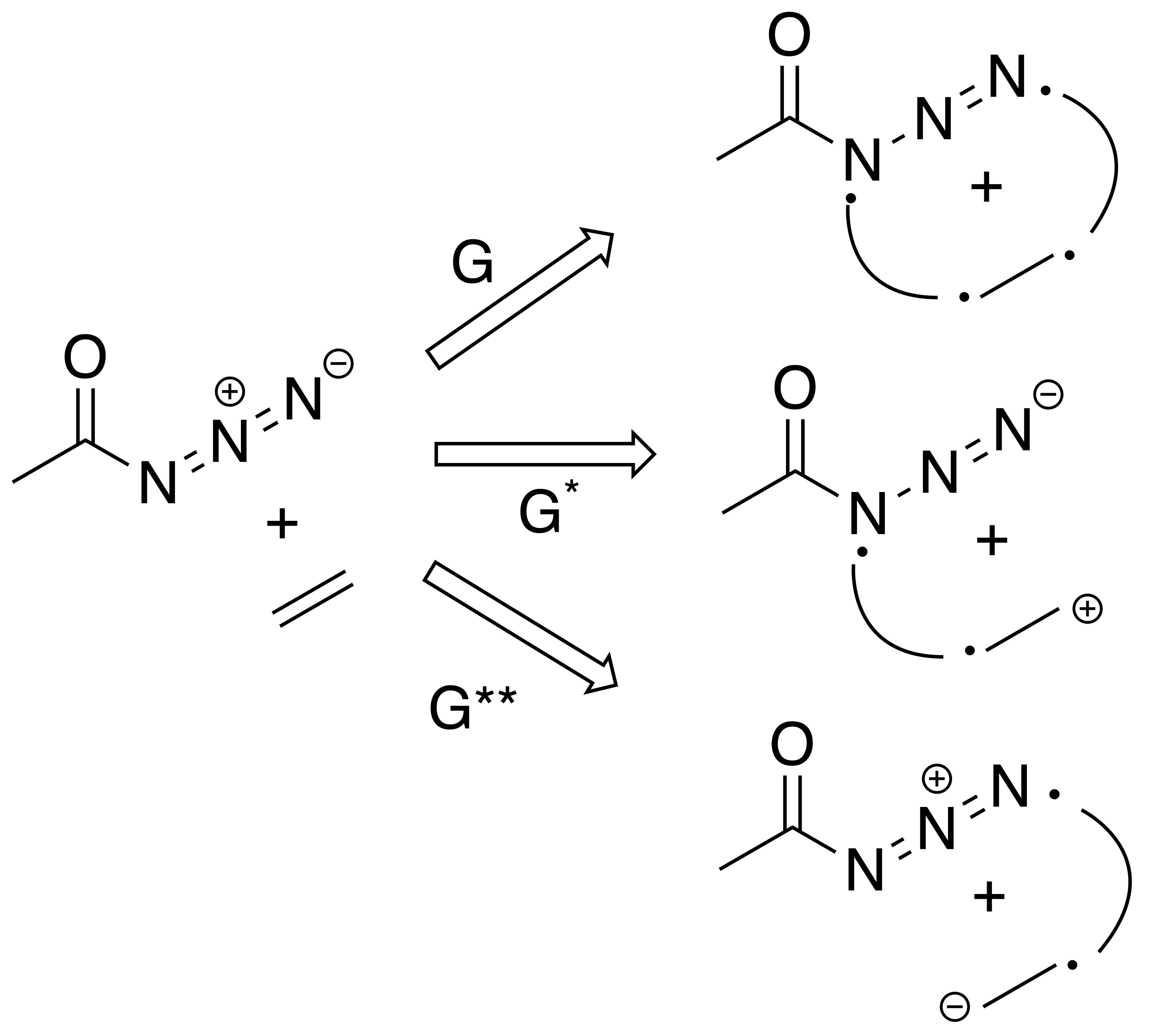}
\caption{(A) Schematic overview of the various promotion gaps that can be defined (cf. Eq. \ref{eq:G}-\ref{eq:G**}).}
\label{fig:promotion}
\end{figure}

The main promotion gap, $G$ is the energy difference between the ground state and the excited state of the reactant system which correlates with the product ground state (Figure \ref{fig:promotion}). In practice, $G$ can be expressed as the sum of the triplet gaps of the reactants,

\begin{equation}\label{eq:G}
G = [{}^{1}E(dipole) - {}^{3}E(dipole)] + [{}^{1}E(dipolarophile) - {}^{3}E(dipolarophile)]
\end{equation}

The other promotion gaps, $G^*$ and $G^{**}$, are the energy differences between the ground state and the charge transfer states in which an electron has been transferred from one reactant to the other and vice versa respectively. 

As such, the following expressions can be defined,

\begin{equation}\label{eq:G*}
G^* = [E(dipole[N-1]) - E(dipole[N])] + [E(dipolarophile[N+1]) - E(dipolarophile[N])]
\end{equation}

\begin{equation}\label{eq:G**}
G^{**} = [E(dipole[N+1]) - E(dipole[N])] + [E(dipolarophile[N-1]) - E(dipolarophile[N])]
\end{equation}

\section{Description of the QM-descriptor pipeline}\label{sec:QM_pipeline}

To generate the selected QM-descriptors for each of the reactant dipoles and dipolarophiles, an automated pipeline for electronic structure calculations was set up. The starting point are the SMILES string representations of the individual molecules. In a first step, 300 conformers are generated and the lowest one in energy is selected with the help of the python package autodE. Within autodE, the ETKDGv2 method is used for the conformer generation and GFN2-xTB is used for subsequent optimization and ranking.

In an second step, single-point DFT calculations (+ NBO population analyses) are performed with Gaussian16 at B3LYP/def2-SVP level-of-theory on the geometries outputted by autodE. The states considered are respectively the lowest singlet and triplet ones and the cationic and anionic ones. Finally, the relevant quantities, needed to construct the descriptors discussed in the previous section, are extracted, and the final descriptors outputted in .pkl-format, with the help of an in-house developed script. 

Overall, the QM-descriptor pipeline is extremely robust, with failure rates of 0.1-0.2\% (calculations for 6 dipoles and 1 dipolarophile failed).

The code associated with this workflow can be accessed from \url{https://github.com/tstuyver/QM_desc_autodE/}.

We want to end this section by noting that we assigned the descriptor values for the uncharged analogs to the three charged biologically inspired fragments featured in the dataset. The reason for doing so is that most of the considered descriptors are associated with electron addition/removal and the presence of charged moieties disjoint from the reacting centers can reasonably be expected to obscure the trends that can otherwise be discerned for these descriptors. Indeed, with the unaltered descriptor values for these 3 molecules, the improvement in accuracy compared to the baseline GNN essentially vanished; modifying the descriptor values leads to the results presented here.

\section{Comparison between the performance of the QM-augmented and regular/baseline GNNs for selective sampling experiments}\label{sec:selective_sampling}

Seven different data splits were considered next to random splitting:

\begin{itemize}
\item All the reactions involving \emph{keto}-based substituents on the negative pole of the dipole, as defined in the corresponding SMILES string, taken as the test set (split 1).
\item All the reactions involving $-CF_3$ substituents taken as the test set (split 2).
\item All the reactions involving $-CN$ substituents on the negative pole of the dipole, as defined in the corresponding SMILES string, taken as the test set (split 3).
\item All the reactions involving $-F$ substituents in the C=C-based dipolarophiles taken as the test set (split 4).
\item All the reactions involving -amide substituents in the C=C-based dipolarophiles taken as the test set (split 5).
\item All the reactions involving -phenyl substituents in the C=C-based dipolarophiles taken as the test set (split 6). 
\item All the reactions for which the dipole core consists of a \emph{CNC}-pattern taken as the test set (split 7).
\end{itemize}

The obtained accuracies for each of these splits, as well as the random splitting baseline (5-fold cross-validation), are shown in Table \ref{tbl:performance_selective} for the respective models. Next to the full QM-augmented GNN, which includes both atom- and reaction-level descriptors, we also considered an analog in which only atom-level descriptors are provided.

From this table, it is clear that the full QM-augmented model reaches a significantly better accuracy than the other models for $G^{\ddagger}$ for 5 out of the 7 splits, as well as for random splitting, with improvements up to 40\% compared to the regular GNN baseline. Even in the 2 splits where the regular GNN analog does perform better, the differences tend to be small. Between the full QM-augmented model and the one that only has access to atom-level descriptors, the differences in accuracy are minor, but still amount to well over 15\% in some cases. This indicates that the inclusion of the reaction-level descriptors is helpful for the model. 

Also for the $G_r$ target, QM-augmentation clearly improves the obtained results, with the regular GNN outperformed in 6 out of 7 splits, sometimes by as much 50-80\%. For this target however, the added benefit of the reaction-level descriptors is less pronounced: in 3 splits, the model with only atom-level descriptors actually performs better than the full QM-augmented one. This is not entirely surprising; the selected reaction-level descriptors, i.e., the promotion gaps, are theoretically connected to reaction barriers, not reaction energies. \cite{shaik2007chemist, stuyver2021promotion}  

\begin{longtable}{cccccc} 
    \caption{Comparison between the performance of the QM-augmented GNN and its regular analog, in terms of mean absolute error (MAE) and root mean square error (RMSE) for the different splits considered.}
  \label{tbl:performance_selective} \\
    \hline \multicolumn{1}{c}{split} & \multicolumn{1}{c}{model type} & \multicolumn{1}{c}{target} & \multicolumn{1}{c}{MAE (kcal/mol} & \multicolumn{1}{c}{RMSE (kcal/mol} \\ \hline 
\endfirsthead
    \hline
    random (5-fold CV) & QM-GNN & \begin{tabular}{@{}c@{}} 
    $G^{\ddagger}$ \\ $G_r$ \end{tabular} & \begin{tabular}{@{}c@{}} \textbf{2.96} \\ \textbf{2.89} \end{tabular} & 
    \begin{tabular}{@{}c@{}} \textbf{4.03} \\ \textbf{4.22} \end{tabular} \\ 
    & QM-GNN (atom desc) & \begin{tabular}{@{}c@{}} 
    $G^{\ddagger}$ \\ $G_r$ \end{tabular} & \begin{tabular}{@{}c@{}} 3.01 \\ 2.92 \end{tabular} & 
    \begin{tabular}{@{}c@{}} 4.09 \\ 4.26 \end{tabular} \\
    & regular GNN & \begin{tabular}{@{}c@{}} 
    $G^{\ddagger}$ \\ $G_r$ \end{tabular} & \begin{tabular}{@{}c@{}} 3.09 \\ 3.09 \end{tabular} & 
    \begin{tabular}{@{}c@{}} 4.17 \\ 4.48 \end{tabular} \\
    \hline

    split 1 & QM-GNN & \begin{tabular}{@{}c@{}} 
    $G^{\ddagger}$ \\ $G_r$ \end{tabular} & \begin{tabular}{@{}c@{}} \textbf{3.95} \\ \textbf{4.58} \end{tabular} & 
    \begin{tabular}{@{}c@{}} \textbf{5.27} \\ \textbf{6.00} \end{tabular} \\
    & QM-GNN (atom desc) & \begin{tabular}{@{}c@{}} 
    $G^{\ddagger}$ \\ $G_r$ \end{tabular} & \begin{tabular}{@{}c@{}} 4.08 \\ 4.89 \end{tabular} & 
    \begin{tabular}{@{}c@{}} 5.47 \\ 6.36  \end{tabular} \\ 
    & regular GNN & \begin{tabular}{@{}c@{}} 
    $G^{\ddagger}$ \\ $G_r$ \end{tabular} & \begin{tabular}{@{}c@{}} 4.68 \\ 6.14 \end{tabular} & 
    \begin{tabular}{@{}c@{}} 6.19 \\ 7.99 \end{tabular} \\ 
    \hline

    split 2 & QM-GNN & \begin{tabular}{@{}c@{}} 
    $G^{\ddagger}$ \\ $G_r$ \end{tabular} & \begin{tabular}{@{}c@{}} \textbf{5.12} \\ 9.27 \end{tabular} & 
    \begin{tabular}{@{}c@{}} \textbf{6.79} \\ 11.73 \end{tabular} \\
    & QM-GNN (atom desc) & \begin{tabular}{@{}c@{}} 
    $G^{\ddagger}$ \\ $G_r$ \end{tabular} & \begin{tabular}{@{}c@{}} 5.28 \\ \textbf{8.40} \end{tabular} & 
    \begin{tabular}{@{}c@{}} 6.90 \\ \textbf{10.94}  \end{tabular} \\ 
    & regular GNN & \begin{tabular}{@{}c@{}} 
    $G^{\ddagger}$ \\ $G_r$ \end{tabular} & \begin{tabular}{@{}c@{}} 7.10 \\ 12.31 \end{tabular} & 
    \begin{tabular}{@{}c@{}} 9.53 \\ 15.53 \end{tabular} \\ 
    \hline

    split 3 & QM-GNN & \begin{tabular}{@{}c@{}} 
    $G^{\ddagger}$ \\ $G_r$ \end{tabular} & \begin{tabular}{@{}c@{}} \textbf{3.40} \\ \textbf{4.43} \end{tabular} & 
    \begin{tabular}{@{}c@{}} \textbf{4.46} \\ \textbf{5.69} \end{tabular} \\
    & QM-GNN (atom desc) & \begin{tabular}{@{}c@{}} 
    $G^{\ddagger}$ \\ $G_r$ \end{tabular} & \begin{tabular}{@{}c@{}} 3.53 \\ 4.82 \end{tabular} & 
    \begin{tabular}{@{}c@{}} 4.65 \\ 6.29  \end{tabular} \\ 
    & regular GNN & \begin{tabular}{@{}c@{}} 
    $G^{\ddagger}$ \\ $G_r$ \end{tabular} & \begin{tabular}{@{}c@{}} 3.92 \\ 5.41 \end{tabular} & 
    \begin{tabular}{@{}c@{}} 5.15 \\ 7.01 \end{tabular} \\ 
    \hline

    split 4 & QM-GNN & \begin{tabular}{@{}c@{}} 
    $G^{\ddagger}$ \\ $G_r$ \end{tabular} & \begin{tabular}{@{}c@{}} 4.38 \\ 8.95 \end{tabular} & 
    \begin{tabular}{@{}c@{}} 5.64 \\ 11.15 \end{tabular} \\
    & QM-GNN (atom desc) & \begin{tabular}{@{}c@{}} 
    $G^{\ddagger}$ \\ $G_r$ \end{tabular} & \begin{tabular}{@{}c@{}} \textbf{4.13} \\ 8.86 \end{tabular} & 
    \begin{tabular}{@{}c@{}} \textbf{5.26} \\ 11.28  \end{tabular} \\ 
    & regular GNN & \begin{tabular}{@{}c@{}} 
    $G^{\ddagger}$ \\ $G_r$ \end{tabular} & \begin{tabular}{@{}c@{}} 4.19 \\ \textbf{7.06} \end{tabular} & 
    \begin{tabular}{@{}c@{}} 5.45 \\ \textbf{9.10} \end{tabular} \\ 
    \hline

    split 5 & QM-GNN & \begin{tabular}{@{}c@{}} 
    $G^{\ddagger}$ \\ $G_r$ \end{tabular} & \begin{tabular}{@{}c@{}} \textbf{4.27} \\ \textbf{4.09} \end{tabular} & 
    \begin{tabular}{@{}c@{}} \textbf{5.45} \\ \textbf{5.22} \end{tabular} \\
    & QM-GNN (atom desc) & \begin{tabular}{@{}c@{}} 
    $G^{\ddagger}$ \\ $G_r$ \end{tabular} & \begin{tabular}{@{}c@{}} 5.02 \\ 4.20 \end{tabular} & 
    \begin{tabular}{@{}c@{}} 6.28 \\ 5.35  \end{tabular} \\ 
    & regular GNN & \begin{tabular}{@{}c@{}} 
    $G^{\ddagger}$ \\ $G_r$ \end{tabular} & \begin{tabular}{@{}c@{}} 5.23 \\ 7.38 \end{tabular} & 
    \begin{tabular}{@{}c@{}} 6.67 \\ 9.06 \end{tabular} \\ 
    \hline

    split 6 & QM-GNN & \begin{tabular}{@{}c@{}} 
    $G^{\ddagger}$ \\ $G_r$ \end{tabular} & \begin{tabular}{@{}c@{}} \textbf{3.92} \\ 5.27 \end{tabular} & 
    \begin{tabular}{@{}c@{}} \textbf{5.11} \\ 7.28 \end{tabular} \\
    & QM-GNN (atom desc) & \begin{tabular}{@{}c@{}} 
    $G^{\ddagger}$ \\ $G_r$ \end{tabular} & \begin{tabular}{@{}c@{}} 4.52 \\ \textbf{4.63} \end{tabular} & 
    \begin{tabular}{@{}c@{}} 5.90 \\ \textbf{6.48}  \end{tabular} \\ 
    & regular GNN & \begin{tabular}{@{}c@{}} 
    $G^{\ddagger}$ \\ $G_r$ \end{tabular} & \begin{tabular}{@{}c@{}} 4.60 \\ 5.71 \end{tabular} & 
    \begin{tabular}{@{}c@{}} 6.05 \\ 7.87 \end{tabular} \\ 
    \hline   

    split 7 & QM-GNN & \begin{tabular}{@{}c@{}} 
    $G^{\ddagger}$ \\ $G_r$ \end{tabular} & \begin{tabular}{@{}c@{}} 7.27 \\ 6.79 \end{tabular} & 
    \begin{tabular}{@{}c@{}} 9.50 \\ \textbf{8.70} \end{tabular} \\
    & QM-GNN (atom desc) & \begin{tabular}{@{}c@{}} 
    $G^{\ddagger}$ \\ $G_r$ \end{tabular} & \begin{tabular}{@{}c@{}} 7.43 \\ \textbf{6.75} \end{tabular} & 
    \begin{tabular}{@{}c@{}} 9.68 \\ \textbf{8.70}  \end{tabular} \\ 
    & regular GNN & \begin{tabular}{@{}c@{}} 
    $G^{\ddagger}$ \\ $G_r$ \end{tabular} & \begin{tabular}{@{}c@{}} \textbf{7.14} \\ 7.39 \end{tabular} & 
    \begin{tabular}{@{}c@{}} \textbf{9.38} \\ 9.69 \end{tabular} \\ 
    \hline    
\end{longtable}

\section{In-sample model performance across active learning iterations}

As discussed in the Methodology section of the main text, during the second iteration of active learning, 2 data points with erroneous $G_r$ values were identified. To facilitate a fair comparison across iterations, we provide two versions of the performance table below: one in which the datasets in iterations 0 and 1 have not been altered (Table \ref{tbl:performance_ensemble}), and one in which the incorrect data points have been retroactively removed (Table \ref{tbl:performance_ensemble_corr}).

\begin{table}
  \caption{Final performance, in terms of mean absolute error (MAE) and root mean square error (RMSE), of the ensemble model constructed from 10 QM-augmented GNNs across active learning iterations. Iteration 0 corresponds to the model trained on the original -- unaltered -- dataset.}
  \label{tbl:performance_ensemble}
  \begin{tabular}{ccccc}
    \hline
    iteration & model type & target & MAE (kcal/mol) & RMSE (kcal/mol) \\
    \hline
    0 & QM-augmented GNN ensemble & \begin{tabular}{@{}c@{}} 
    $G^{\ddagger}$ \\ $G_r$ \end{tabular} & \begin{tabular}{@{}c@{}} 2.64 \\ 2.40 \end{tabular} & 
    \begin{tabular}{@{}c@{}} 3.63 \\ 3.64 \end{tabular} \\ 
    & regular GNN ensemble & \begin{tabular}{@{}c@{}} 
    $G^{\ddagger}$ \\ $G_r$ \end{tabular} & \begin{tabular}{@{}c@{}} 2.76 \\ 2.52 \end{tabular} & 
    \begin{tabular}{@{}c@{}} 3.77 \\ 3.80 \end{tabular} \\ 
    \hline
    
    1 & QM-augmented GNN ensemble & \begin{tabular}{@{}c@{}} 
    $G^{\ddagger}$ \\ $G_r$ \end{tabular} & \begin{tabular}{@{}c@{}} 2.56 \\ 2.35 \end{tabular} & 
    \begin{tabular}{@{}c@{}} 3.55 \\ 3.58 \end{tabular} \\ 
    & regular GNN ensemble & \begin{tabular}{@{}c@{}} 
    $G^{\ddagger}$ \\ $G_r$ \end{tabular} & \begin{tabular}{@{}c@{}} 2.69 \\ 2.47 \end{tabular} & 
    \begin{tabular}{@{}c@{}} 3.68 \\ 3.68 \end{tabular} \\ 
    \hline
    
    2 & QM-augmented GNN ensemble & \begin{tabular}{@{}c@{}} 
    $G^{\ddagger}$ \\ $G_r$ \end{tabular} & \begin{tabular}{@{}c@{}} 2.49 \\ 2.22 \end{tabular} & 
    \begin{tabular}{@{}c@{}} 3.45 \\ 3.17 \end{tabular} \\ 
    & regular GNN ensemble & \begin{tabular}{@{}c@{}} 
    $G^{\ddagger}$ \\ $G_r$ \end{tabular} & \begin{tabular}{@{}c@{}} 2.62 \\ 2.35 \end{tabular} & 
    \begin{tabular}{@{}c@{}} 3.60 \\ 3.29 \end{tabular} \\ 
    \hline

    3 & QM-augmented GNN ensemble & \begin{tabular}{@{}c@{}} 
    $G^{\ddagger}$ \\ $G_r$ \end{tabular} & \begin{tabular}{@{}c@{}} 2.47 \\ 2.21 \end{tabular} & 
    \begin{tabular}{@{}c@{}} 3.43 \\ 3.16 \end{tabular} \\ 
    & regular GNN ensemble & \begin{tabular}{@{}c@{}} 
    $G^{\ddagger}$ \\ $G_r$ \end{tabular} & \begin{tabular}{@{}c@{}} 2.60 \\ 2.33 \end{tabular} & 
    \begin{tabular}{@{}c@{}} 3.58 \\ 3.29 \end{tabular} \\ 
  \end{tabular}
\end{table}

\begin{table}
  \caption{Final performance, in terms of mean absolute error (MAE) and root mean square error (RMSE), of the ensemble model constructed from 10 QM-augmented GNNs across active learning iterations (retroactively corrected datasets for iteration 0 and 1).}
  \label{tbl:performance_ensemble_corr}
  \begin{tabular}{ccccc}
    \hline
    iteration & model type & target & MAE (kcal/mol) & RMSE (kcal/mol) \\
    \hline
    0 & QM-augmented GNN ensemble & \begin{tabular}{@{}c@{}} 
    $G^{\ddagger}$ \\ $G_r$ \end{tabular} & \begin{tabular}{@{}c@{}} 2.64 \\ 2.33 \end{tabular} & 
    \begin{tabular}{@{}c@{}} 3.64 \\ 3.24 \end{tabular} \\ 
    & regular GNN ensemble & \begin{tabular}{@{}c@{}} 
    $G^{\ddagger}$ \\ $G_r$ \end{tabular} & \begin{tabular}{@{}c@{}} 2.76 \\ 2.46 \end{tabular} & 
    \begin{tabular}{@{}c@{}} 3.77 \\ 3.41 \end{tabular} \\ 
    \hline
    
    1 & QM-augmented GNN ensemble & \begin{tabular}{@{}c@{}} 
    $G^{\ddagger}$ \\ $G_r$ \end{tabular} & \begin{tabular}{@{}c@{}} 2.55 \\ 2.30 \end{tabular} & 
    \begin{tabular}{@{}c@{}} 3.52 \\ 3.23 \end{tabular} \\ 
    & regular GNN ensemble & \begin{tabular}{@{}c@{}} 
    $G^{\ddagger}$ \\ $G_r$ \end{tabular} & \begin{tabular}{@{}c@{}} 2.66 \\ 2.42 \end{tabular} & 
    \begin{tabular}{@{}c@{}} 3.65 \\ 3.36 \end{tabular} \\ 
    \hline
    
    2 & QM-augmented GNN ensemble & \begin{tabular}{@{}c@{}} 
    $G^{\ddagger}$ \\ $G_r$ \end{tabular} & \begin{tabular}{@{}c@{}} 2.49 \\ 2.22 \end{tabular} & 
    \begin{tabular}{@{}c@{}} 3.45 \\ 3.17 \end{tabular} \\ 
    & regular GNN ensemble & \begin{tabular}{@{}c@{}} 
    $G^{\ddagger}$ \\ $G_r$ \end{tabular} & \begin{tabular}{@{}c@{}} 2.62 \\ 2.35 \end{tabular} & 
    \begin{tabular}{@{}c@{}} 3.60 \\ 3.29 \end{tabular} \\ 
    \hline

    3 & QM-augmented GNN ensemble & \begin{tabular}{@{}c@{}} 
    $G^{\ddagger}$ \\ $G_r$ \end{tabular} & \begin{tabular}{@{}c@{}} 2.47 \\ 2.21 \end{tabular} & 
    \begin{tabular}{@{}c@{}} 3.43 \\ 3.16 \end{tabular} \\ 
    & regular GNN ensemble & \begin{tabular}{@{}c@{}} 
    $G^{\ddagger}$ \\ $G_r$ \end{tabular} & \begin{tabular}{@{}c@{}} 2.60 \\ 2.33 \end{tabular} & 
    \begin{tabular}{@{}c@{}} 3.58 \\ 3.29 \end{tabular} \\ 
    
  \end{tabular}
\end{table}

\section{Validation performance across active learning iterations}

\begin{table}
  \caption{Performance on the selected, out-of-sample, validation reactions, in terms of mean absolute error (MAE) and root mean square error (RMSE), of the ensemble model constructed from 10 QM-augmented GNNs across active learning iterations. Iteration 0 corresponds to the model trained on the original dataset.}
  \label{tbl:performance_ensemble_validation}
  \begin{tabular}{ccccc}
    \hline
    iteration & model type & target & MAE (kcal/mol) & RMSE (kcal/mol) \\
    \hline
    0 & QM-augmented GNN ensemble & \begin{tabular}{@{}c@{}} 
    $G^{\ddagger}$ \\ $G_r$ \end{tabular} & \begin{tabular}{@{}c@{}} 2.54 \\ 2.51 \end{tabular} & 
    \begin{tabular}{@{}c@{}} 3.52 \\ 3.51 \end{tabular} \\ 
    \hline
    1 & QM-augmented GNN ensemble & \begin{tabular}{@{}c@{}} 
    $G^{\ddagger}$ \\ $G_r$ \end{tabular} & \begin{tabular}{@{}c@{}} 2.36 \\ 2.46 \end{tabular} & 
    \begin{tabular}{@{}c@{}} 3.21 \\ 3.88 \end{tabular} \\ 
    \hline
    2 & QM-augmented GNN ensemble & \begin{tabular}{@{}c@{}} 
    $G^{\ddagger}$ \\ $G_r$ \end{tabular} & \begin{tabular}{@{}c@{}} 2.69 \\ 2.39 \end{tabular} & 
    \begin{tabular}{@{}c@{}} 3.85 \\ 3.43 \end{tabular} \\ 
  \end{tabular}
\end{table}

\section{In-depth discussion of the active learning results}\label{sec:active_learning_results}

\subsection{Active learning - iteration 0}
In the first training round, only 488 reactions passed all the filtering steps, involving 40 distinct dipoles and 111 distinct dipolarophiles. The dipole represented cover examples from the entire chemical space, i.e., various scaffolds from the allyl-type, propargyl/allenyl-type and cyclic type dipoles can be identified, whereas the dipoloraphiles selected mainly originate from the cyclooctyne scaffold (over 90\%), with no examples from either the oxo-norbornadiene or norbornene scaffold. 20 of the retained dipoles were selected, and up to 5 synthetic reactions were sampled for each. The 76 reactions obtained in this manner were complemented with all the corresponding reactions involving biologically-inspired motifs (cf. Figure 1), resulting in a final list of 533 validation reactions. These reactions were subsequently computed with our computational workflow, yielding 473 reaction profiles (failure rate of 11.3\%). 443 of these reactions did not yet feature in the original dataset, and hence, those were included for retraining of the model in iteration 1. 

Comparing the computed activation and reaction energies with the predictions made by the model indicates that excellent accuracy on the -- out-of-sample -- validation reactions is achieved (Table \ref{tbl:performance_ensemble_validation}). In fact, the obtained mean absolute errors (MAE) and root mean square errors (RMSE) are slightly better than what is obtained during cross-validation of the original dataset (\emph{vide supra}). This can be attributed in part to differences in the respective data point distributions: while in the original dataset, the proportion of synthetic reactions amounted to over 50\%, 80\% of the validation reactions involve biologically inspired fragments (which involve less diversity and hence were already sampled more densely in the original dataset). Support for this hypothesis is found by computing the errors separately for the biofragment and synthetic reactions; the MAE for the latter amounts to 3.07 kcal/mol, whereas for the former it amounts to only 2.32 kcal/mol.

Despite the excellent accuracy, fully resolving promising and non-promising bio-orthogonal click reactions is challenging for the model trained on the original dataset: upon explicit calculation, less than half of the 20 dipoles sampled (9) for validation exhibit no barriers below 24 kcal/mol with any of the biologically inspired fragments. For the validated dipoles, we identify 20 synthetic reactions exhibiting an activation energy well below 22 kcal/mol in total, i.e., they exhibit all the criteria to be plausible bio-orthogonal click reaction candidates -- at least equally promising as the prototypical methyl and acyl azides discussed above. 

\subsection{Active learning - iteration 1}

Retraining our model with the validated reactions included only marginally improves the accuracy of the predictions overall (cf. Table \ref{tbl:performance_ensemble_validation}), but it makes the model markedly less conservative at the margins: the number of synthetic reactions retained after the filtering steps jumps to 975 now. Remarkably, the number of distinct synthetic dipolarophiles represented in the retained reactions increases almost by a factor of 2, from 111 to 163. It should be noted that the vast majority of the retained dipolarophiles are still cyclooctyne based (142), and the oxo-norbornadiene and norbornene scaffolds are still barely represented (5 and 0 respectively). For the dipoles, on the other hand, we observe a reduction in diversity; only 25 are retained now, 12 of which did not pass all the filtering steps during the initial round. 

The seemingly contradictory trends, increasing dipolarophile and decreasing dipole diversity, can be reconciled straightforwardly: the increased diversity in the training set during (re)training, causes a fairly consistent -- yet moderate -- downward shift in the predicted activation energies for click-like reactions. Indeed, for the selected synthetic reactions for which predictions were available also in iteration 0, the activation energy is on average 2.2 kcal/mol lower in energy, which is enough for many reactions to cross the imposed activation energy threshold. As a consequence, more dipoles get discarded due to excessive reactivity with a biologically inspired fragment. At the same time, many new reactions are retained since the dipolarophile becomes fast enough for bio-orthogonal click applications according to our model.  

In an attempt to sample more diverse reactions (only cyclooctyne-based synthetic reactions had been sampled for validation during the initial round, and the dominance of this scaffold did not decrease upon retraining), we modified the sampling strategy for validation in the first iteration. Now, we selected every dipole that had not yet been sampled during the initial round and sampled all the non-cyclooctyne-based synthetic reactions first; the number of synthetic reactions was then supplemented with additional cyclooctyne-based reactions until at least 5 were selected in total. Taking this approach, 139 synthetic reactions got sampled, out of which over half (87) did not include a dipolarophile with a cyclooctyne scaffold, resulting in a final list of 551 validation reactions. Passing each of these data points through our computational workflow, 464 reaction profiles were obtained (failure rate of 15.8\%). 445 of these reactions did not yet feature in the dataset, and hence they were included for retraining in iteration 2.

Remarkably, comparing the computed activation and reaction energies with the predicted values reveals that the model is now able to resolve promising and non-promising bio-orthogonal click reactions much better than the initial model: 83\% of the sampled dipoles (13 out 16) exhibit no barriers below 24 kcal/mol with any of the biologically inspired fragments, and 84 out of 128 successfully computed -- and not previously considered -- candidate reactions pass all the filtering steps to be plausible bio-orthogonal click reactions.

\subsection{Active learning - iteration 2}

In the second iteration, \footnote{It should be noted here that during the analysis of this validation batch, 2 reaction profiles were identified where autodE incorrectly computed the product to be more stable than the TS by a significant margin. Further analysis revealed that this erroneous result was caused by incorrect conformer selection by autodE. Additionally, we identified 2 data points in the original dataset for which this was also the case (see main text). All 4 data points were removed from the dataset for the remainder of the study.} the number of synthetic reactions retained stabilizes at 903. This time, 27 distinct dipoles make the cut, 11 of which were not retained in either iteration 0 or 1. Additionally, 178 unique dipolarophiles passed all filtering steps, 157 of which are cyclooctyne-based dipolarophiles. 

In contrast to the previous iteration, no uniform shift in the predicted activation energies is observed: on average, the predictions for the selected synthetic reactions in this round have shifted downward by a negligible 0.1 kcal/mol compared to iteration 1. 

Since only 11 dipoles retained in this iteration had not been validated before, we decided to sample all of them. For each dipole, up to 10 synthetic reactions were selected. Preference was given to non-cyclooctyne-based dipolarophiles and then cyclooctyne-based dipolarophiles which had not been computed before. If the target of 10 synthetic reactions was not reached in this manner; additional -- previously sampled -- cyclooctyne reactions were included as well. In total, this yielded 44 reactions, of which 42 with a cyclooctyne scaffold, and 2 with an oxo-norbornadiene scaffold. Complementing this synthetic reaction list with the corresponding reactions involving biologically inspired motifs, a final list of 296 validation reactions was obtained. Due to the fact that up to this point, no reactions involving a norbornene-based dipolarophile were sampled, we added to this list the 10 synthetic reactions involving norbornene with the lowest predicted activation energy. Out of the resulting 306 validation reactions, 276 yielded a reaction profile (failure rate of 9.8\%). 258 of these reactions did not yet feature in the dataset, and hence they were
included for the final retraining (\emph{vide infra}). 

During this iteration, 8 out of 11 dipoles sampled exhibit no barrier below 24 kcal/mol with any of the biologically inspired fragments, and 16 out of 42 successfully computed -- and not previously considered -- candidate reactions pass all the filtering steps to be plausible bio-orthogonal click reactions. Note that the validation rate of this iteration lies well below the rate obtained for iteration 1 (presumably because the most promising dipoles were sampled in this round), yet still significantly above the rate obtained for iteration 0. 

For the extra norbornene reactions sampled during this iteration, we observe that the predictions for the activation energy made by the model are poor on average (MAE of 3.6 kcal/mol and RMSE of almost 5 kcal/mol). Notably, the predictions for this target are all clustered in a narrow range (24-26 kcal/mol), whereas the computed activation energies cover a much broader one from 18 to 37 kcal/mol. Part of the reason for the poor quality of the predictions for this quantity presumably lies in the limited diversity of the sampling of this specific dipolarophile scaffold in the original dataset -- the model will usually only make predictions within the range it has encountered during training -- underscoring the importance of additional (active learning) sampling. For the reaction energies, the problem outlined above is much less pronounced. Here, the predictions are in fact excellent, with an MAE of only 1.5 kcal/mol and an RMSE of 1.9 kcal/mol. Despite the poor quality of the activation energy predictions for this scaffold, it is important to note that none of the extra sampled norbornene-based synthetic reactions would have come anywhere close to passing all the generic filtering steps outlined above.

Overall, the results for this validation round suggest that the added value of additional sampling has started to diminish. This is corroborated by retraining the model one last time; the number of retained reactions now rises slightly to 1123, with 30 unique dipoles and 221 unique dipolarophiles. From these 30 dipoles, 24 have previously been sampled, and also more than 80\% of the retained dipolarophiles (184) have made it into the list of retained reactions before. As expected, the activation energies predicted for the norbornene scaffold exhibit much more diversity now (predicted activation energies of 20-21 kcal/mol are obtained now), but still, none of them pass all filtering steps. Putting everything together, we decided to halt our active learning procedure at this point and proceed with the final retrained model.

\putbib[biblio_ms]
\end{bibunit}

\end{document}